\documentclass[notitlepage,letterpaper,aps,twocolumn,amsmath,amsfonts,nofootinbib,preprintnumbers,superscriptaddress,eqsecnum,secnumarabic]{revtex4-1}
\pdfoutput = 1
\usepackage{natbib}
\usepackage{amssymb,amsmath,latexsym,mathrsfs}
\usepackage{bm}
\usepackage{url}
\usepackage{epsfig,graphicx,verbatim,xspace,multirow,mathtools}
\usepackage{array}
\usepackage{enumitem}
\usepackage{graphicx,slashed}
\usepackage[usenames,dvipsnames]{color}
\usepackage{todonotes}
\usepackage{color}
\usepackage{chngcntr}

\counterwithout{equation}{section} 
\usepackage[normalem]{ulem}
\usepackage[breaklinks,colorlinks,urlcolor=blue,citecolor=blue,linkcolor=blue]{hyperref}
\usepackage{epsfig,verbatim,xspace,multirow,mathtools}
\usepackage{subfiles}

\newcommand{\Eq}[1]{Eq.~\eqref{#1}}
\newcommand{\Fig}[1]{Fig.~\ref{#1}}
\newcommand{\Tab}[1]{Tab.~\ref{#1}}
\newcommand{\Sec}[1]{Section~\ref{#1}}
\newcommand{\App}[1]{Appendix \ref{#1}}
\newcommand{\ie}{\emph{i.e.~}}
\newcommand{\eg}{\emph{e.g.~}}

\begin{document}

\preprint{\tt TUM-HEP 1318/21}

\title{The Hubble Tension as a Hint of Leptogenesis and Neutrino Mass Generation}

\author{Miguel Escudero}
\email{miguel.escudero@tum.de}
\thanks{ORCID: \href{https://orcid.org/0000-0002-4487-8742}{0000-0002-4487-8742}}
\affiliation{Physik-Department, Technische Universit{\"{a}}t, M{\"{u}}nchen, James-Franck-Stra{\ss}e, 85748 Garching, Germany}
\author{Samuel J. Witte}
\email{s.j.witte@uva.nl}
\thanks{ORCID: \href{https://orcid.org/0000-0003-4649-3085}{0000-0003-4649-3085}}
\affiliation{Gravitation Astroparticle Physics Amsterdam (GRAPPA), Institute for Theoretical Physics Amsterdam and Delta Institute for Theoretical Physics, University of Amsterdam, Science Park 904, 1098 XH Amsterdam, The Netherlands}
\begin{abstract}
\noindent 
The majoron, a neutrinophilic pseudo-Goldstone boson conventionally arising in the context of neutrino mass models, can damp neutrino free-streaming and inject additional energy density into neutrinos prior to recombination. The combination of these effects for an eV-scale mass majoron has been shown to ameliorate the outstanding $H_0$ tension, however only if one introduces additional dark radiation at the level of $\Delta N_{\rm eff} \sim 0.5$. We show here that models of low-scale leptogenesis can naturally source this dark radiation by generating a primordial population of majorons from the decays of GeV-scale sterile neutrinos in the early Universe. Using a posterior predictive distribution conditioned on Planck2018+BAO data, we show that the value of $H_0$ observed by the SH$_0$ES collaboration is expected to occur at the level of $\sim 10\%$ in the primordial majoron cosmology (to be compared with $\sim 0.1\%$ in the case of $\Lambda$CDM). This insight provides an intriguing connection between the neutrino mass mechanism, the baryon asymmetry of the Universe, and the discrepant measurements of $H_0$.\end{abstract}

\maketitle

\section{Introduction}
\vspace{0.2cm}

\noindent \textbf{The Hubble Tension, circa early 2021.} Despite its simplicity, the standard cosmological model (\ie $\Lambda$CDM) has proven to be remarkably successful in describing the vast array of cosmological observations at hand. However in recent years, a growing discrepancy has emerged between the value of $H_0$ as predicted by $\Lambda$CDM~\cite{planck,Aghanim:2019ame,Abbott:2017smn,Addison:2017fdm,Schoneberg:2019wmt,Cuceu:2019for}, $H_0 \simeq 67.4\pm0.5\,\text{km}/\text{s}/\text{Mpc}$ and local observations that favor a significantly larger value, with a central values ranging between $70 \lesssim H_0 \lesssim 74 \,\text{km}/\text{s}/\text{Mpc}$, coming from \eg type Ia supernovae~\cite{Riess:2016jrr,Riess:2019cxk,Dhawan:2017ywl,Burns:2018ggj,Riess:2018uxu,Freedman:2019jwv,Yuan:2019npk,Reid:2019tiq,Riess:2020fzl}, strong gravitational lensing~\cite{Bonvin:2016crt,Birrer:2018vtm,Rusu:2019xrq,Chen:2019ejq,Wong:2019kwg}, surface brightness fluctuations~\cite{Blakeslee:2021rqi}, and megamasers~\cite{Reid:2008nm} (see \eg\cite{Verde:2019ivm,Riess:2020sih} for recent reviews). This tension has now reached a significance quantified at the $4-6 \,\sigma$ level~\cite{Riess:2020sih}, and appears across an array of different datasets with seemingly independent systematics. It is thus necessary to now consider the very real possibility that this discrepancy is arising from a failure of the $\Lambda$CDM model to accurately describe the evolution of the Universe.

\vspace{0.1cm}

In this light, a large number of potential solutions have been proposed which typically fall into one of two categories: those which modify the Universe at late times ($z \lesssim 1$) and those which modify the dynamics and evolution near recombination ($10^3 \, \lesssim z \lesssim 10^5$) -- we refer the reader to~\cite{Mortsell:2018mfj,Poulin:2018zxs,Nunes:2018xbm,DiValentino:2019ffd,Vattis:2019efj,Li:2019san,Visinelli:2019qqu,Yang:2021egn} and to~\cite{Poulin:2018cxd,Agrawal:2019lmo,Lin:2019qug,Smith:2019ihp,DEramo:2018vss,Escudero:2019gzq,Bringmann:2018jpr,Pandey:2019plg,Blinov:2020uvz,Blinov:2020hmc,Lancaster:2017ksf,Kreisch:2019yzn,Park:2019ibn,Sakstein:2019fmf,Archidiacono:2020yey,Choi:2020tqp,Freese:2021rjq} for several recent proposals of each type, and to~\cite{diValentino:rev} for a recent comprehensive review of models. Cosmological observations from type Ia supernovae and baryonic acoustic oscillations (BAO) severely limit the viability of late-time solutions (see \eg\cite{Bernal:2016gxb,Aylor:2018drw,Knox:2019rjx}). In order for early Universe solutions to be successful in raising the inferred value of $H_0$, one must modify the expansion rate near recombination. While these models often struggle to maintain the high-quality fit to CMB data that is obtained in $\Lambda$CDM, this classification of solutions has proven thus far to be most successful in producing agreement between all cosmological datasets -- see~\cite{Knox:2019rjx}. Finding meaningful ways to motivate the novel physics required, however, has been challenging -- many of the proposed solutions require complicated and un-motivated models which are fine-tuned in order to ensure that the effects `turn-on' at the correct epoch and produce sufficiently sizable shift in $H_0$.

\begin{figure*}
	\includegraphics[width=\textwidth]{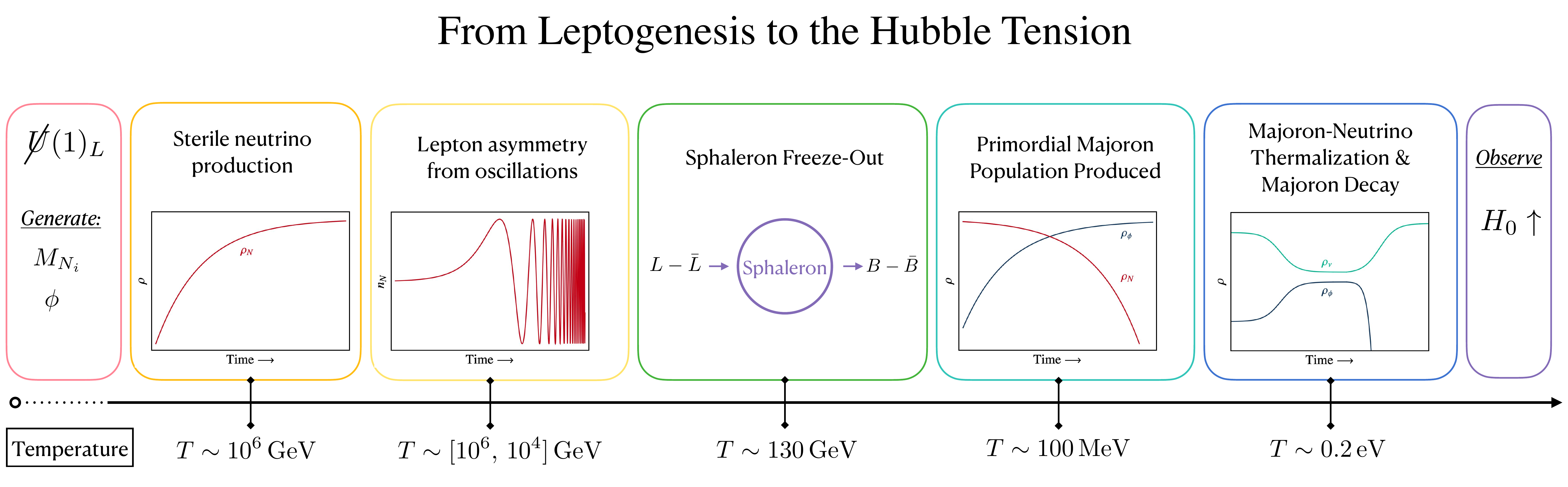}
	\caption{Cosmological timeline illustrating the connection between low-scale leptogenesis and the majoron solution to the Hubble tension. At early times (high temperatures), a global $U(1)_{L}$ symmetry is spontaneously broken, generating sterile neutrino masses and giving rise to a pseudo-Goldstone boson: the majoron ($\phi$). Sterile neutrinos start to be sizeably produced (but do not equilibrate) at $T \sim 10^6$ GeV. Then, at $T \sim [10^6-10^4]$ GeV the CP violating oscillations of these sterile neutrinos generate a net primordial lepton asymmetry in the Standard Model. Soon after the electroweak phase transition (at $T \sim 130$ GeV) sphalerons freeze-out and yield a final baryon asymmetry from the initial lepton asymmetry. After sphaleron freeze-out, sterile neutrinos and majorons thermalize with the plasma, and later decouple when sterile neutrinos decay. In particular, for $\sim$ GeV scale sterile neutrinos this occurs at temperatures below the QCD phase transition $T \lesssim 100\,\text{MeV}$. Finally, right before recombination, majorons with $m_\phi \sim 1\,\text{eV}$ re-thermalize with active neutrinos ($\bar{\nu}\nu \to \phi$) before decaying  ($ \phi \to \bar{\nu}\nu$), generating a larger inferred cosmological value of $H_0$.}\label{fig:cartoon}
\end{figure*}

\vspace{0.1cm}

\noindent \textbf{The Hubble tension and a light Majoron.} Recently, in Ref.~\cite{Escudero:2019gvw} (see also~\cite{Escudero:2020hkf} for a short summary), the authors illustrated that a light majoron, naturally arising in neutrino mass models from the spontaneous breaking of a global lepton number symmetry~\cite{Chikashige:1980ui,Gelmini:1980re,Georgi:1981pg,Schechter:1981cv}, could partially counteract the effect of additional dark radiation (\ie $\Delta N_{\rm eff}$), pushing the inferred value of $H_0$ to larger values while maintaining a good fit to the CMB data. The presence of the majoron has two effects~\cite{Chacko:2003dt}. First, the interactions damp the free streaming nature of neutrinos, which in turn damps the anisotropic stress; since the anisotropic stress sources the metric, the net effect is a time-dependent modification to the growth of potential wells~\cite{Bashinsky:2003tk}. Second, for large enough interactions, the majorons can thermalize with neutrinos between Big Bang Nucleosynthesis (BBN) and recombination. Upon becoming non-relativistic  these majorons decay back into neutrinos, producing a net enhancement in $\Delta N_{\rm eff}$~\cite{Chacko:2003dt,Escudero:2019gvw}. Ensuring the effects of the majoron occur at the correct epoch requires a mass $m_\phi$ near the eV scale; this scale, however, is not arbitrary, and can easily be motivated should global lepton number be explicitly broken by physics at the Planck scale~\cite{Rothstein:1992rh,Akhmedov:1992hi}, as might be expected in theories of quantum gravity~\cite{Kallosh:1995hi,Banks:2010zn,Witten:2017hdv,Harlow:2018jwu,Fichet:2019ugl,Alvey:2020nyh}. In order to generate a sufficient level of damping one requires neutrino-majoron couplings $\lambda \sim 10^{-13}$~\cite{Escudero:2019gvw,Escudero:2020hkf}\footnote{The model discussed here has, on occasion, been confused with that of the strongly interaction neutrino solution proposed in~\cite{Kreisch:2019yzn,Park:2019ibn}. In light of this, we take the opportunity here to highlight the many differences. First, the solution of~\cite{Kreisch:2019yzn,Park:2019ibn} requires a neutrino self-interaction cross section 10 orders of magnitude larger than that present in the Standard Model. This, in turn, requires a new $ {\rm MeV}$-scale neutrinophilic boson with order one couplings. These values are not motivated in neutrino mass models, and are robustly excluded by experimental data unless the boson interacts only with $\tau$ neutrinos~\cite{Blinov:2019gcj,Lyu:2020lps,Brdar:2020nbj}. Next, the solution requires an additional contribution of $\Delta N_{\rm eff} \sim 1$, a value robustly excluded by BBN~\cite{Pitrou:2018cgg,Fields:2019pfx} -- see also~\cite{Huang:2021dba,Seto:2021xua} for a recent assessment of the BBN bounds and~\cite{Berbig:2020wve,He:2020zns} for models trying to evade these constraints. Finally, the observed shift in $H_0$ only occurs when polarization data is not included in the fit~\cite{Kreisch:2019yzn,Choudhury:2020tka,Das:2020xke,Brinckmann:2020bcn}, while the results for the majoron model discussed here are robust to the inclusion of this dataset. Thus, while the proposed models both involve neutrinophilic bosons, they are in fact remarkably different.}. This coupling, when interpreted in the context of the type-I seesaw favors a lepton symmetry breaking scale slightly above the electroweak scale ($v_L \sim 1\,\text{TeV}$). Arguably,  the only unmotivated aspect of this proposed solution is the apparent {\it ad hoc} contribution of $\Delta N_{\rm eff}$, preferring values $\sim 0.5$, which are in mild tension with BBN~\cite{Pitrou:2018cgg,Fields:2019pfx}.

\vspace{0.3cm}

\textbf{Primordial Majorons from Leptogenesis.} In this work we attempt to source the additional dark radiation required to resolve the $H_0$ tension from a primordial population of majorons. We show explicitly that these particles can be produced from the decays of GeV-scale sterile neutrinos in the early Universe. Coincidentally, sterile neutrinos at the GeV scale are precisely those required for a successful implementation of low-scale leptogenesis via sterile neutrino oscillations, \ie ARS leptogenesis~\cite{Akhmedov:1998qx} (see also~\cite{Asaka:2005pn,Shaposhnikov:2008pf,Drewes:2017zyw}). We verify explicitly that symmetry breaking scales $v_L \sim (0.01-1)\,\text{TeV}$ required to resolve the Hubble tension can be made fully consistent with conventional ARS leptogenesis, so long as the Higgs mixing is small enough so as to avoid thermalizing the scalar responsible for breaking lepton number, and that the lepton number phase transition occurs at $T>10^{4}-10^6\,\text{GeV}$. The scenario proposed here thus offers an intriguing connection between the $H_0$ tension, the neutrino mass mechanism, and the generation of the baryon asymmetry of the Universe. Fig.~\ref{fig:cartoon} shows a sketch of the thermal history, highlighting the main ingredients of our proposal.

This manuscript is organized as follows. We begin by introducing the well-known singlet majoron model in \Sec{sec:model}. In \Sec{sec:ars} we first discuss the requirements in order to successfully produce the baryon asymmetry of the Universe via the ARS leptogenesis mechanism, and then compute the  thermal evolution and subsequent decays of the sterile neutrinos responsible for sourcing the primordial majoron abundance. \Sec{sec:cosmo} describes the cosmological evolution of the majoron-neutrino system, and presents the results of a MCMC performed using {\tt Planck2018 + BAO} data. We present a summary and our conclusions in \Sec{sec:con}. We finish in \Sec{sec:outlook} by discussing some interesting avenues for future work, and we refer the reader to the Appendices for various technical details.

\section{The Singlet Majoron Model}\label{sec:model}

Throughout this manuscript we work with the well-known singlet majoron model~\cite{Chikashige:1980ui} (see also~\cite{Schechter:1981cv}) in which Majorana masses for the right-handed neutrinos $N_R$ are generated from the spontaneous breaking of a global $U(1)_L$ lepton number symmetry\footnote{Global symmetries are also expected to be explicitly broken by quantum gravity~\cite{Kallosh:1995hi,Banks:2010zn,Witten:2017hdv,Harlow:2018jwu,Fichet:2019ugl,Alvey:2020nyh}. This has two important consequences. First, this breaking will generate a mass for the majoron; if the breaking is perturbative, one can show that dimension 5 Planck-suppressed operators naturally generate masses in the range $1 {\rm eV} \lesssim m_\phi \lesssim 100$ keV~\cite{Rothstein:1992rh,Akhmedov:1992hi}, which coincidentally overlaps with the range of interest for the Hubble tension (see~\cite{Escudero:2019gvw} for a discussion). Additionally, an explicit breaking of the symmetry guarantees that topological defects will naturally decay on short timescales~\cite{Vilenkin:1982ks,Kawasaki:2013ae}, and thus  consequently pose no threat of over-closing the Universe.}. In this set-up, the small neutrino masses then arise from the type-I seesaw mechanism~\cite{Minkowski:1977sc,Mohapatra:1980yp,GellMann:1980vs,Yanagida:1980xy,Schechter:1980gr}. 

This model is realized by augmenting the Standard Model (SM) with $n \geq 2$ right handed neutrinos with lepton number $L =+1$ and a scalar field $\Phi$ with $L = +2$, all singlets under the SM gauge group. This particle content and charge arrangements, together with the requirement of renormalizability of the interactions, leads to the following Lagrangian
\begin{align}\label{eq:Lagrangian}
\mathcal{L}  \supset &\, (\partial_\mu \Phi)^\dagger (\partial^\mu \Phi) - V_\Phi + i\bar{N}_{Ri}\gamma^\mu \partial_\mu N_{Ri} \nonumber \\
 &-\frac{\lambda_{N_{ij}}}{\sqrt{2}} \Phi \, \overline{N}_{R, \, i} N_{R, \, j}^c - h_{\alpha i} \overline {L}_L^\alpha H N_{Ri} +\text{h.c.}  \, ,
\end{align}
 where $\lambda_{N_{ij}}$ are the $\Phi$-$N$ Yukawa couplings, and $h_{\alpha i}$ are the Higgs-Lepton-$N$ Yukawa couplings. The scalar potential $V_\Phi$ is given by
\begin{align}
 V_\Phi= -\mu_\Phi^2 \Phi^\dagger \Phi + \lambda_\Phi (\Phi^\dagger \Phi)^2 - 
\lambda_{\Phi H} (H^\dagger H) \, (\Phi^\dagger \Phi)\,.
\end{align}
Upon spontaneous symmetry breaking (SSB) of the $U(1)_L$ symmetry, the scalar $\Phi$  acquires a vacuum expectation value $v_L$ and will generate Majorana masses for the sterile neutrinos $M_N = \lambda_N v_L$. Since $U(1)_L$ is a global symmetry, a pseudo-Goldstone boson appears on the spectrum: the \textit{majoron} $\phi$~\cite{Chikashige:1980ui}. After SSB, it is convenient to parametrize $\Phi$ as 
\begin{equation}\label{eq:V_phi}
	\Phi = \frac{v_L + \rho}{\sqrt{2}}e^{i {\phi}/{v_L}} \, ,
\end{equation}
where  $\rho$ is a CP even scalar,  which in the limit $\lambda_{\Phi H} \rightarrow 0$ will have a tree level mass given by $m_\rho^2 = 2\lambda_\Phi v_L^2$.

Only after the SSB of the electroweak symmetry will Dirac neutrino masses appear, $\left[m_D\right]_{\alpha i} = h_{\alpha i} v_H/\sqrt{2}$ with $v_H = 246\,\text{GeV}$. Diagonalizing the neutrino mass matrix in the limit $m_D \ll M_N$ yields light active Majorana neutrinos with masses of the order:
\begin{align}\label{eq:seesawmass}
m_\nu \simeq m_D^2 /M_N\,.
\end{align}
Assuming no strong cancellations occur, one typically expects the heavy sterile neutrinos to have very small mixings with light active states (see~\cite{Casas:2001sr} for a general case), with a magnitude roughly given by
\begin{align}\label{eq:seesawmixing}
|\theta|^2 \simeq m_\nu/M_N \simeq 5\times10^{-11} \frac{m_\nu}{0.05\,\text{eV}}\frac{1\,\text{GeV}}{M_N}\,,
\end{align}
and with Higgs' Yukawa couplings of the order
\begin{align}\label{eq:seesaw_DiracY}
\!\! |h | \simeq  \frac{\sqrt{2 m_\nu M_N}}{v_H} \simeq  4\times 10^{-8} \sqrt{ \frac{m_\nu}{0.05\,\text{eV}}}\sqrt{\frac{{M_N}}{1\,\text{GeV}}}\,.
 \end{align}

One of the conventional appeals of the seesaw mechanism is that the fine-tuning of the Yukawa couplings required to generate the active neutrino masses can be ameliorated when $M_N > v_H$. Successful low-scale leptogenesis, however, requires GeV-scale sterile neutrino masses, which are capable of reducing, but not removing the aforementioned fine-tuning problem (see \Eq{eq:seesaw_DiracY}). While this naively appears to remove at least part of the original appeal, it is worth emphasizing that there exist mixing textures (\ie non-trivial $h_{\alpha i}$)\footnote{In this context, we refer to~\cite{Arias-Aragon:2020qip} for a recent embedding of our set up within the minimal lepton flavor violating framework.} which can substantially enhance $|\theta|^2$ relative to the value quoted in \Eq{eq:seesawmixing} (and further remove any need for tuning), however for the sake of simplicity and concreteness we choose to work within the prototypical seesaw limit.

\vspace{0.15cm}
\begin{center} \textbf{\small A. Neutrino-Majoron Interactions}  \end{center}
\vspace{0.1cm}

With Eq.~\eqref{eq:Lagrangian} in hand, one can enumerate the novel interactions that arise between the neutrinos (both active and sterile), the majoron, and $\rho$. Working in the mass basis of $N$ and $\nu$, the relevant interactions in the seesaw limit are~\cite{Pilaftsis:1991ug,Pilaftsis:1993af}
\begin{align}\label{eq:Lag_massbasis}
	\mathcal{L} \supset & - \frac{\lambda_{N}}{2} \, \left[ \rho \bar{N} N - i \phi  \bar{N} \gamma_5 N   \right] \, \nonumber ,\\
				            & - \frac{\lambda_{N\nu}}{2} \, \left[ i \rho ( \bar{N}\gamma_5 \nu + \bar{\nu} \gamma_5 N) - \phi ( \bar{N} \nu +\bar{\nu}N )  \right] \, \nonumber ,\\
	                                     & + \frac{\lambda_{\nu}}{2} \, \left[ \rho \bar{\nu} \nu - i \phi  \bar{\nu} \gamma_5 \nu   \right] \, ,
\end{align}
where the couplings are given by\footnote{Notice that Eq.~\eqref{eq:lambda_def} is smaller by a factor of 2 with respect to the one used in~\cite{Escudero:2019gvw}. We are grateful to Manuel Masip for pointing out this error.}
\begin{align}\label{eq:Lag_massbasis_coup}
	\lambda_N & =  \frac{M_N}{v_L} = 10^{-3} \frac{M_N}{1\,\text{GeV}} \frac{1\,\text{TeV}}{v_L} \,,\\
	\lambda_{N\nu} & =  \frac{\sqrt{m_\nu M_N}}{v_L} = 7\times 10^{-9} \sqrt{\frac{M_N}{1\,\text{GeV}}} \frac{1\,\text{TeV}}{v_L} \,,\\
	\lambda_\nu & =  \frac{m_\nu}{v_L} = 5\times10^{-14} \frac{m_\nu}{0.05\,\text{eV}} \frac{1\,\text{TeV}}{v_L} \label{eq:lambda_def} \,.
\end{align}
Here we have omitted generation indices, however this is valid when considering the three light neutrino mass eigenstates as $\lambda_\nu$ is a diagonal matrix up to tiny $\mathcal{O}(|\theta|^2)$ corrections~\cite{Schechter:1981cv}. It is worth highlighting that interactions between the majoron and charged fermions are both neutrino mass and loop-suppressed, and are thus expected to appear at the level of $\lambda_{\phi ee} \lesssim 10^{-20}$~\cite{Chikashige:1980ui}. The majoron is thus, for all intents, a truly neutrinophilic boson. 

There are two interactions in particular that are worth mentioning at this point. In order to generate a sizable primordial population of majorons in the very early Universe ($T> 10\,\text{MeV}$), we must require sterile neutrinos to efficiently decay into majorons, and in order to ensure majorons thermalize with neutrinos at late-times, we must ensure the inverse decay of active neutrinos to majorons exceeds the Hubble rate near recombination. As we shall see, both of these requirements are satisfied for values of $M_N \sim \text{GeV}$ and $v_L \sim v_H$. 

Let us begin by assuming sterile neutrinos have thermalized at temperatures $T \gtrsim M_N$, and identify the condition necessary to generate a sizable branching fraction to majorons. 
The decay rate of sterile neutrinos into a majoron and active neutrino is
\begin{align}
\Gamma(N\to \nu \phi) = \frac{m_\nu}{16\pi} \left(\frac{M_N}{v_L}\right)^2\, .
\end{align}
This should be compared with the decay rate into SM particles, which for sterile neutrino masses $0.1\,\text{GeV} < M_N < 10\,\text{GeV}$ is roughly given by $\Gamma(N\to {\rm SM}) \sim 10\times \Gamma(N\to 3 \nu)$  (see e.g.~\cite{GonzalezGarcia:1990fb}), where
\begin{align}
\Gamma(N\to 3\nu) =  \frac{1}{12}   \frac{m_\nu}{32\pi} \left(\frac{M_N}{v_H}\right)^4  \,.
\end{align}
Therefore, the ratio between the decay rates is
 \begin{align}\label{eq:ratio}
\frac{\Gamma(N\to \nu \phi)}{\Gamma(N\to \text{SM})} \simeq 8\times 10^{3} \left(\frac{1\,\text{GeV}}{M_N}\right)^2 \left(\frac{1\,\text{TeV}}{v_L}\right)^2\,,
\end{align}
implying a branching ratio $\text{Br}(N\to \nu\phi) \simeq 1$ for all relevant parameter space. 

We will show explicitly in the next section that in order for the majoron solution to remain viable (at least in the seesaw limit), we require $v_L$ to be $\lesssim 2$ TeV and $M_N \lesssim 3$ GeV; the former is necessary in order to ensure neutrino-majoron interactions can damp the neutrino anisotropic stress, and the latter is required in order to generate a sufficiently large contribution to $\Delta N_{\rm eff}$ (if sterile neutrinos decay at earlier times the energy density of any majoron population produced will be diluted by subsequent entropy dumps). \Eq{eq:ratio} then shows that sterile neutrino decays will be efficient in generating majorons across the entirety of the parameter space of interest.

Now let us turn our attention to the late-time phenomenology, where we must require majorons to thermalize with active neutrinos near recombination. For the small couplings and masses of interest ($\lambda \sim 10^{-13}\,, m_\phi \sim 1\,\text{eV}$), this process proceeds via the inverse decays of active neutrinos ($\bar{\nu} \nu \leftrightarrow  \phi$). The efficiency of this process is governed by the majoron decay rate into $\bar{\nu}\nu$:
\begin{align}
\Gamma_\phi  = \frac{\lambda_\nu^2}{16\pi} m_\phi \, \sqrt{1-\frac{4m_\nu^2}{m_\phi^2}} \simeq \frac{\lambda_\nu^2}{16\pi} m_\phi \, .
\end{align}
Notice that the interaction strength in this model is directly proportional to the light neutrino masses. Neutrino oscillation measurements imply~\cite{Esteban:2020cvm,deSalas:2020pgw}: $\sqrt{m_{2}^2-m_1^2} \simeq 0.0086$ eV and that $\sqrt{m_3^2-m_1^2} \simeq  0.05\,\text{eV}$ or $\sqrt{m_2^2-m_3^2} \simeq 0.05\,\text{eV}$, depending of whether the neutrino mass ordering is normal or inverted, respectively. In \cite{Escudero:2019gvw}, the authors considered the scenario in which the majoron interacted equally with all three neutrinos -- which given the observed mass splittings is realized when $\sum m_\nu \gtrsim 0.15\,\text{eV}$. The two alternative limiting cases are realized when the lightest neutrino is approximately massless. In normal ordering this will correspond to $m_1 \sim 0$, and then majorons interact almost exclusively with the most massive eigenstate $m_3$ (\ie the number of interacting neutrinos is $N_{\rm int} = 1$). On the other hand, in inverted ordering, $m_3 = 0$ and then majorons interact with the two, nearly degenerate, neutrino eigenstates $1$ and $2$ (\ie $N_{\rm int} = 2$).  While it is far from obvious, we will show in what follows that cosmological observables are not strongly sensitive to the difference in $N_{\rm int}$, provided of course that $N_{\rm int }\geq 1$.

For convenience in what follows, we will define here an effective width parameter
\begin{align}\label{eq:gamma_eff2}
\Gamma_{\rm eff} \equiv \left(\frac{\lambda_\nu}{4\times 10^{-14}} \right)^2 \, \left(\frac{0.1 \, {\rm eV}}{m_\phi} \right) \, ,
\end{align}
where the normalization has been chosen such that for $\Gamma_{\rm eff} \gtrsim 1$ majorons thermalize with active neutrinos via inverse neutrino decays. This parameter is thus more intimately connected with the cosmological implications of majorons at late-times than the direct coupling $\lambda_\nu$ itself.

\section{$\Delta N_{\rm eff}$ as a product of Leptogenesis}\label{sec:ars}

\noindent The majoron solution to the $H_0$ tension, presented by the authors in~\cite{Escudero:2019gvw}, requires three novel ingredients: \textit{(1)} a majoron with mass $m_\phi \sim {\rm eV}$ (which could be generated by an explicit perturbative breaking of the $U(1)_L$ symmetry via dimension 5 Planck-scale suppressed operators~\cite{Rothstein:1992rh,Akhmedov:1992hi}),  \textit{(2)}  a coupling to active neutrinos at the level $\lambda \sim 10^{-13}$ (corresponding to $v_L \sim v_H$), and \textit{(3)} an additional contribution to  $\Delta N_{\rm eff} \sim 0.5$. In this section we show that the novel contribution to $\Delta N_{\rm eff}$, which had previously been introduced in an {\emph{ad hoc}} manner, can be  sourced directly from a primordial population of majorons. This occurs naturally if the masses of the sterile neutrinos are roughly $M_N \sim \text{GeV}$, which is coincidentally exactly the mass scale required for a successful implementation of ARS leptogenesis. 

Here, we begin by outlining the main ingredients of the ARS leptogenesis framework, focusing in particular on whether any feature of the singlet majoron model could prevent or inhibit the generation of the baryon asymmetry of the Universe. We then argue that sterile neutrinos in the ARS leptogenesis framework inevitably lead to a thermal population of majorons after the electroweak phase transition, which subsequently decouples as sterile neutrinos decay, yielding a sizable majoron primordial abundance as relevant for CMB and BBN observations. 

\vspace{0.15cm}
\begin{center} \textbf{\small A. ARS Leptogenesis and the Majoron}  \end{center}
\vspace{0.1cm}

The idea behind ARS leptogenesis is as follows~\cite{Akhmedov:1998qx} (see also~\cite{Asaka:2005pn,Shaposhnikov:2008pf}, and~\cite{Drewes:2017zyw} for a review): 
\begin{enumerate}
\item One assumes that at sufficiently high temperatures (\ie after reheating) there are no sterile neutrinos in the plasma. At temperatures $T\gg T_{\rm EW} \sim 160 \, {\rm GeV}$ an out-of-equilibrium population of sterile neutrinos is slowly produced via the small Dirac Yukawa couplings $h_{\alpha i}$.
\item Having been produced, these sterile neutrinos will undergo efficient CP-violating oscillations when $t_{\rm osc} \sim 1/H$. For degenerate sterile neutrinos, this corresponds to temperatures 
\begin{align}\label{eq:TLepto}
 \qquad T_{\rm lepto} \sim   10^{5}\,\text{GeV} \left(\frac{M_N}{10\,\text{GeV}}\right)^{2/3} \left(\frac{\Delta M}{10^{-4}\,M_N}\right)^{1/3}
\end{align}
where $\Delta M$ is the mass difference between a pair of sterile neutrinos. 
\item These oscillations will generate lepton asymmetries in each of the sterile neutrinos individually, but in such a way that the total lepton number asymmetry is still zero. However, sphaleron processes will only convert the SM lepton asymmetries into a baryonic one. Then, when sphalerons freeze out at $T \sim 130\,\text{GeV}$, the baryon asymmetry present at that temperature is frozen and remains constant until today. Thus, as long as one sterile neutrino has not thermalized by $T_{\rm EW}$ (such that there are non-vanishing SM leptonic asymmetries), a non-zero baryon asymmetry will have been generated.
\end{enumerate}
Clearly, the combination of these steps meet the three Sakharov conditions and allow for successful baryogenesis. Rigorous calculations for the case of two sterile neutrinos have shown that the baryon asymmetry of the Universe can be successfully generated in the context of the seesaw limit for $0.1\,\text{GeV}\lesssim M_N \lesssim 10\,\text{GeV}$ and $\Delta M/M_N \sim 10^{-7}-10^{-5}$, see e.g.~\cite{Hernandez:2015wna,Hernandez:2016kel,Eijima:2018qke,Klaric:2020lov}. For the case of three sterile neutrinos, similar calculations have shown that successful baryogenesis can be achieved without such strong mass degeneracy~\cite{Drewes:2012ma}.

Importantly, ARS leptogenesis has {\emph{not}} yet been rigorously investigated in the context of the singlet majoron model. To our knowledge, the only reference to have discussed this issue focused on identifying a minimal, but model-dependent, set of requirements for successful leptogenesis~\cite{Caputo:2018zky}. In order to ensure that the majoron model of interest here can indeed generate the observed baryon asymmetry of the Universe, we revisit the requirements identified in \cite{Caputo:2018zky} using a relaxed set of assumptions.

There are three key requirements in order to maintain the efficiency of the ARS mechanism within the majoron model. Firstly, sterile neutrinos cannot thermalize with the Standard Model plasma at temperatures $T > T_{\rm EW}$, otherwise the lepton asymmetry (and thus also the baryon asymmetry) will vanish. Secondly, the sterile neutrinos must undergo CP-violating oscillations. Thirdly, such oscillations must be coherent at $T_{\rm lepto}$ because it is then when the primordial lepton asymmetry is generated. 

The thermalization of sterile neutrinos can occur via processes of the type $\phi \phi \to \bar{N}N$, $\rho \rho \to \bar{N}N$ and $\rho \to N \bar{N}$. If either of these scalar states have thermalized at high temperatures, then avoiding thermalization of the sterile neutrinos amounts to requiring small sterile neutrino couplings, or equivalently large vevs. In particular, one finds $v_L > 10^{5}-10^6\,\text{GeV}$~\cite{Caputo:2018zky}. Obviously this is in conflict with the requirement for majorons to re-thermalize with neutrinos near recombination, which requires $v_L \lesssim 2$ TeV. It is reasonable, however, to expect that these states would not have thermalized at early times since they are inherently a part of the sterile neutrino sector -- which in ARS leptogenesis are assumed not to be produced during reheating. Should that be the case, we have verified explicitly that none of the processes mentioned above will generate a thermal sterile neutrino population at $T > T_{\rm EW}$, provided that $|\lambda_{\Phi H}| < 10^{-7}$ (see Appendix~\ref{sec:ARS_app}).

In order for the CP violating oscillations of sterile neutrinos to be effective in the early Universe, the $U(1)_L$ symmetry should be broken at $T_c > T_{\rm lepto} \sim (10^4-10^6)\,\text{GeV}$.  In the majoron model the sterile neutrino mass is a time dependent parameter controlled by the vacuum expectation value of the $\Phi$ field: $M_N(T) = \lambda_N \sqrt{2} \left< \Phi \right>(T) $, therefore one must also ensure that the $U(1)_L$ symmetry is spontaneously broken at $T > T_{\rm lepto}$ (see Eq.~\eqref{eq:TLepto}). By studying the 1-loop thermal corrections to the $U(1)_L$ potential, see Appendix~\ref{sec:ARS_app} for the details, we have shown that the condition  $T_c > T_{\rm lepto} \sim (10^4-10^6)\,\text{GeV}$ can be translated into a bound on the Higgs-scalar mixing at the level of:
\begin{align}\label{eq:condition_ARS}
|\lambda_{\Phi H}| <  4.6 \times 10^{-7}\, \frac{v_L}{1\,\text{TeV}} \sqrt{\frac{10^5\,\text{GeV}}{T_c}} \,.
\end{align} 

Finally, within the ARS mechanism, the primordial lepton asymmetry is mainly generated when sterile neutrinos start to oscillate at $T_{\rm lepto} \sim 10^5\,\text{GeV}$. In this stage, it is key that the coherence of such oscillations is maintained. In particular, processes of the type $\phi N \leftrightarrow \phi N$ should not be faster than the oscillation rate $\Gamma \simeq \Delta M^2/4E $ at $T\sim T_{\rm lepto}$. Explicitly, see Appendix~\ref{sec:ARS_app}, this requirement can be translated into a rather weak bound on the $\lambda_N$ coupling given by:
\begin{align}\label{eq:condition_ARS_lambdaN}
\lambda_N  = \frac{M_N}{v_L}  < 0.07 \,  \sqrt{\frac{T_{\rm Lept}}{10^5\,\text{GeV}}} \,  \sqrt{\frac{4\times 10^{-8}}{|h |}} \,,
\end{align} 
which we can clearly appreciate implies a rather mild hierarchy between $M_N$ and $v_L$.

Collectively, these conditions imply that ARS leptogenesis in the context of the singlet majoron model will likely remain unaltered, so long as the Higgs-portal coupling is sufficiently small ($|\lambda_{\Phi H}|  \lesssim 10^{-7}$). While this requirement may appear at first sight to be tuned, we would like to point out that the smallness of this coupling is maintained by quantum corrections. Using \texttt{SARAH}~\cite{Staub:2008uz,Staub:2015kfa}, we have calculated the two-loop beta function for the running of  $\lambda_{\Phi H}$, which has contributions in the form \textit{i)} $d \lambda_{\Phi H}/d\log\mu \propto \lambda_{\Phi H}$ and \textit{ii)} $d \lambda_{\Phi H}/d\log\mu \propto h_N^2 \lambda_N^2 $. The former of these is inherently small, and the latter is suppressed by the active-sterile neutrino mixings. Thus, the requirement $|\lambda_{\Phi H}|  \lesssim 10^{-7}$ is stable under radiative corrections.

\begin{figure}
	\includegraphics[width=0.48\textwidth]{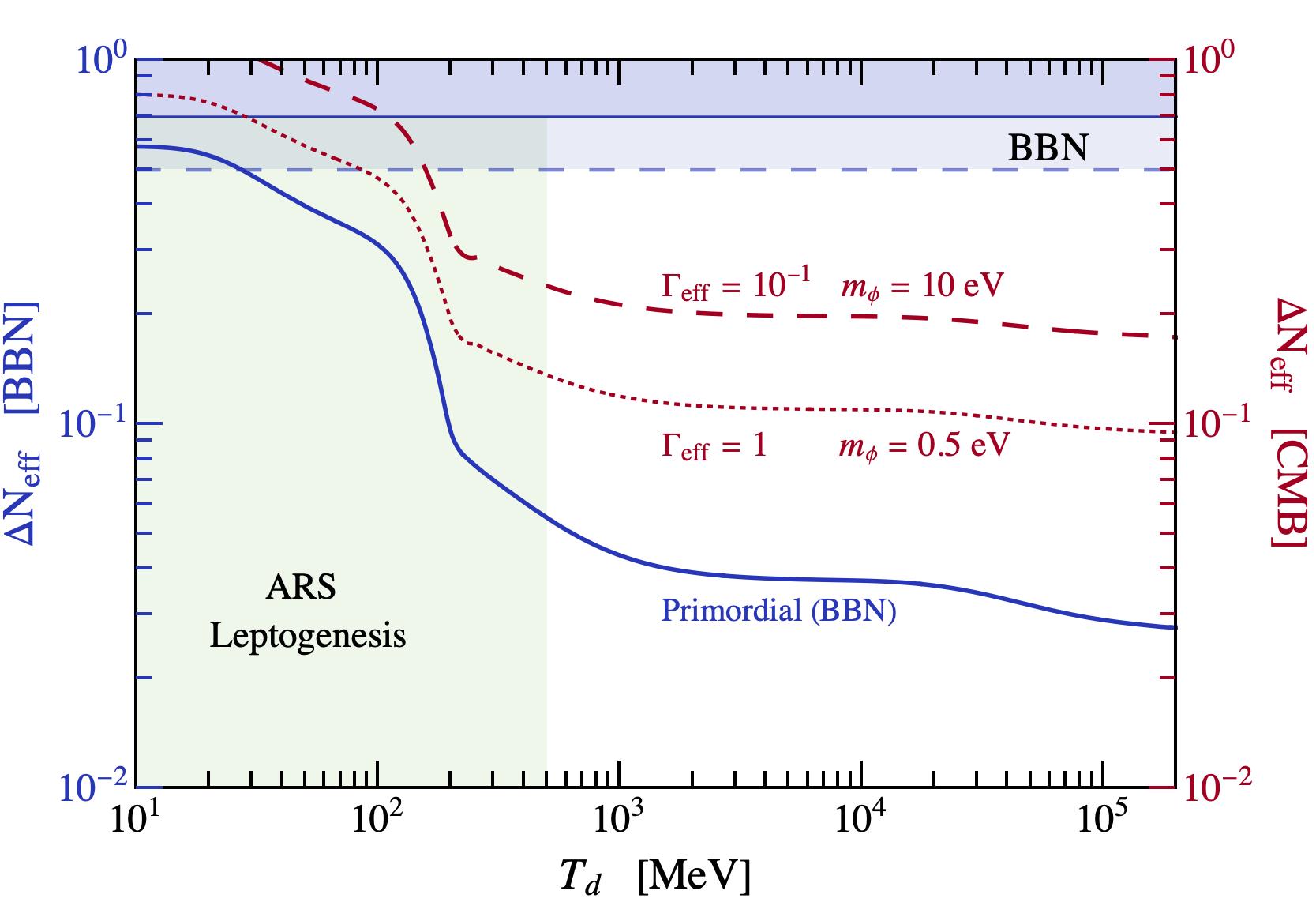}
	\caption{Contribution from a primordial majoron population to $\Delta N_{\rm eff}$ at BBN (blue), assuming $m_\phi \ll 1$ MeV, as a function of decoupling temperature $T_d$ (obtained using the entropy density in the SM from~\cite{Laine:2015kra}). For majorons with $\Gamma_{\rm eff} \lesssim 1$ and $m_\phi \gtrsim 0.1$ eV, the contribution to $\Delta N_{\rm eff}$ at recombination (red) can be greatly enhanced due to the fact that majorons do not decay immediately after becoming relativistic; this point is illustrated for $\Gamma_{\rm eff} = 10^{-1}$ and $m_\phi = 10$ (dashed). In green we show the approximate parameter space relevant for ARS leptogenesis, see Eq.~\eqref{eq:Td_maintext}. We show two constraints from BBN (dark and light blue horizontal regions; to be compared with blue solid line) which arise from two different determinations of $Y_p$~\cite{Izotov:2014fga,Aver:2020fon}, see Appendix~\ref{app:bbn}. }\label{fig:neff} 
\end{figure}

\newpage 
\vspace{0.35cm}
\begin{center} \textbf{\small B. Primordial Majoron Population from \\ Sterile Neutrino Decays}  \end{center}
\vspace{0.20cm}

Successful ARS leptogenesis requires sterile neutrino masses $0.1\,\text{GeV}\lesssim M_N \lesssim 10\,\text{GeV}$, which are generically expected to thermalize with the SM plasma at temperatures $1\,\text{GeV}\lesssim T \lesssim 80\,\text{GeV}$~\cite{Ghiglieri:2016xye}. For the sterile neutrino masses and vevs of interest, sterile neutrino annihilations will be efficient in thermalizing a majoron population during this epoch (note that $\Gamma(\bar{N}N \leftrightarrow \phi \phi)/H \sim 300\,(M_N/\text{GeV})^2 \left(2\,\text{TeV}/v_L\right)^4 $ at $T \sim M_N$). Eventually, as the sterile neutrinos decay, the thermalized majoron population will decouple from the plasma and freeze out while relativistic. Comparing the decay rate to the Hubble expansion we can estimate the temperature at which majorons decouple (see Appendix~\ref{sec:PrimordialMaj_app} for details):  
\begin{align}\label{eq:Td_maintext}
T_d  &\lesssim \left(\frac{M_N}{13}\right) \,,
\end{align} 
which holds for sterile neutrinos with $v_L \lesssim v_H$.

In the event that majoron decoupling is instantaneous, one can compute the energy density stored in the majoron population at BBN by simply accounting for the entropy dilution in the SM plasma after decoupling -- this is shown in \Fig{fig:neff} (blue line labeled `Primordial') as a function of decoupling temperature $T_d$. Here, we express the energy density in terms of  $\Delta N_{\rm eff} \equiv N_{\rm eff}-N_{\rm eff}^{\rm SM}$, where
\begin{equation}
	N_{\rm eff}\equiv \frac{8}{7}\,\left(\frac{11}{4}\right)^{4/3}\left(\frac{\rho_{\rm rad}-\rho_\gamma}{\rho_\gamma} \right)\,,
\end{equation}
and $N_{\rm eff}^{\rm SM}=3.044$~\cite{Escudero:2020dfa,Akita:2020szl,Froustey:2020mcq,Bennett:2020zkv,Hansen:2020vgm}.  

\vspace{0.15cm}

It is worth noting that very light particles which decouple at extremely late times near $T_d \sim 10$ MeV may be in tension with constraints from Big Bang Nucleosynthesis (BBN). There is, however, some ambiguity as to where these constraints truly lie; the reason being that the leading local cosmology-independent determinations of the primordial helium abundance $Y_p$ rely on spectroscopic observations of HII in metal poor galaxies, which could suffer from systematics. The most recent estimate of $Y_P$ is \cite{Aver:2020fon}: $Y_P = 0.2453\pm 0.0034$, but there are other independent analyses that report substantially larger values~\cite{Izotov:2014fga}: $Y_P = 0.2551 \pm 0.0022$. In order to account for the possibility of additional systematics in the determination of $Y_p$, we present throughout constraints on $\Delta N_{\rm eff}$ from these two distinct analyses -- we defer a more detailed discussion of the BBN systematics and the sensitivity to $\Delta N_{\rm eff}$ to Appendix~\ref{app:bbn}. These constraints are plotted in \Fig{fig:neff} using horizontal blue lines (the excluded regions are shaded in blue, and differentiated using solid and dashed lines).

\vspace{0.15cm}
\newpage

\begin{figure}
	\includegraphics[width=0.46\textwidth]{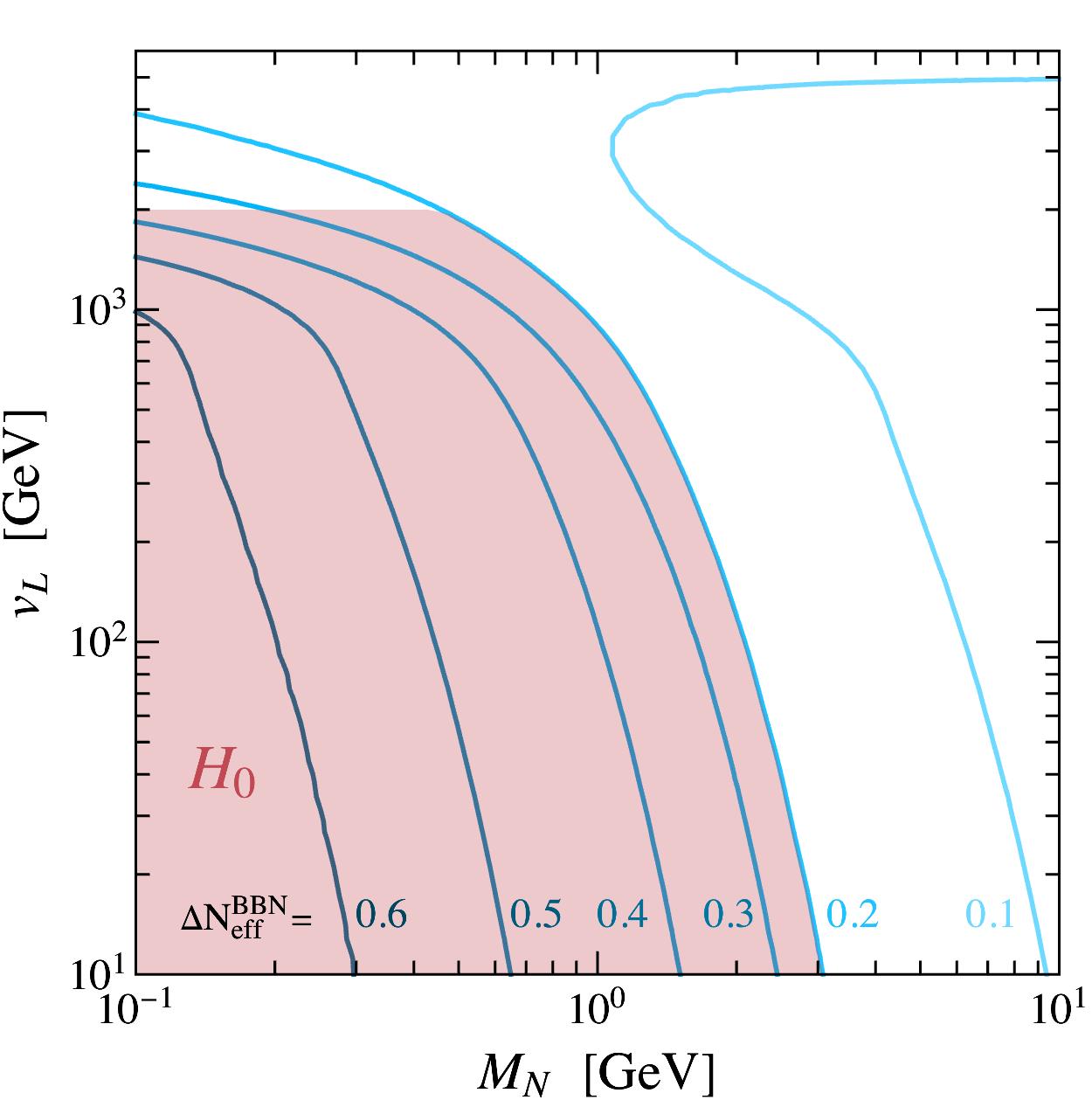}
	\caption{$\Delta N_{\rm eff}$ at BBN as a function of the sterile neutrino mass $M_N$ and the scale of lepton number breaking $v_L$. We have highlighted in red the approximate region preferred to solve the Hubble tension (defined here as the region for which $\Delta N_{\rm eff} \gtrsim 0.2$ and $v_L \leq 2$ TeV). Various $\Delta N_{\rm eff}$ contours are highlighted in blue, including the approximate fiducial model adopted later in this work which roughly corresponds to $\Delta N_{\rm eff} \sim 0.4$. Note that BBN constrains imply $\Delta N_{\rm eff}^{\rm BBN} \lesssim [0.5, 0.7]$, with the uncertainty reflecting different measurements of the primordial Helium abundance, see Appendix~\ref{app:bbn}. } \label{fig:neff_BBN} 
\end{figure}

Should sterile neutrinos decay at $T\lesssim T_{\rm QCD} \sim 200\,\text{MeV}$, the effect of non-instantaneous majoron decoupling can lead to significant changes in the expected energy density of the primordial population. The reason here being simply that the SM plasma undergoes stronger and more rapid entropy dumps arising from the QCD phase transition. In order to more properly estimate the energy density in the parameter space of interest we have solved for the thermodynamic evolution of the $N$ and $\phi$ number densities in the early Universe, and evolved this system to temperatures $T \sim 1\,\text{MeV}$ for a wide array of sterile neutrino masses and interaction strengths. This was done using the methods of~\cite{Escudero:2018mvt,Escudero:2020dfa}, in which one assumes each population can be approximately described by a thermal distribution with a time-dependent temperature and chemical potential. Following this approach, we show the contribution to $\Delta N_{\rm eff}$ as a function of $M_N$ and $v_L$ in \Fig{fig:neff_BBN}. We also highlight the preferred parameter space for the $H_0$ tension (red region). The details of this analysis are contained in Appendix~\ref{sec:PrimordialMaj_app}.

\section{Majoron Cosmology}\label{sec:cosmo}
\vspace{-0.1cm}

Having shown in the previous section that the same sterile neutrinos responsible for leptogenesis can also generate a sizable primordial population of majorons, we now turn our attention to the cosmological evolution of majorons after BBN. The story here is more straightforward: majorons with $\Gamma_{\rm eff} \gtrsim 1$ thermalize with neutrinos at temperatures $T \sim m_\phi$ via $\bar{\nu}\nu \leftrightarrow \phi$ processes, and subsequently decay into neutrinos after becoming non-relativistic a short time later. There are two effects of this process: \textit{(1)} a further increase in $\Delta N_{\rm eff}$~\cite{Escudero:2019gvw,Chacko:2003dt} and \textit{(2)} a damping of the neutrino free streaming~\cite{Bashinsky:2003tk,Chacko:2003dt}. The former point has been illustrated in \Fig{fig:neff}, where the contribution to $\Delta N_{\rm eff}$ at CMB (red) is compared for two different choices of parameters to the contribution to $\Delta N_{\rm eff}$ at BBN (blue). Assuming the majoron decays prior to recombination, models with $\Gamma_{\rm eff} \ll 1$ lead to larger energy densities, as weakly coupled majorons remain non-relativistic for longer times before decaying. While the effect of damping neutrino free streaming is more subtle, these interactions can lead a noticeable imprint on the CMB multipoles. We outline our treatment of both of these effects in this section, and discuss the results of performing MCMCs to current cosmological data.

\vspace{0.15cm}
\begin{center} \textbf{\small A. Background Evolution}  \end{center}
\vspace{0.1cm}

In order to describe the background evolution of majorons and neutrinos in the early Universe we once again use the formalism developed in~\cite{Escudero:2018mvt,Escudero:2020dfa} (described above). This was shown in~\cite{Escudero:2020dfa} to provide a highly accurate description of late-time thermalization of neutrinophilic bosons. We defer the details of these calculations to Appendix~\ref{app:majorons_CMB}. 

\begin{figure}
	\includegraphics[width=0.48\textwidth]{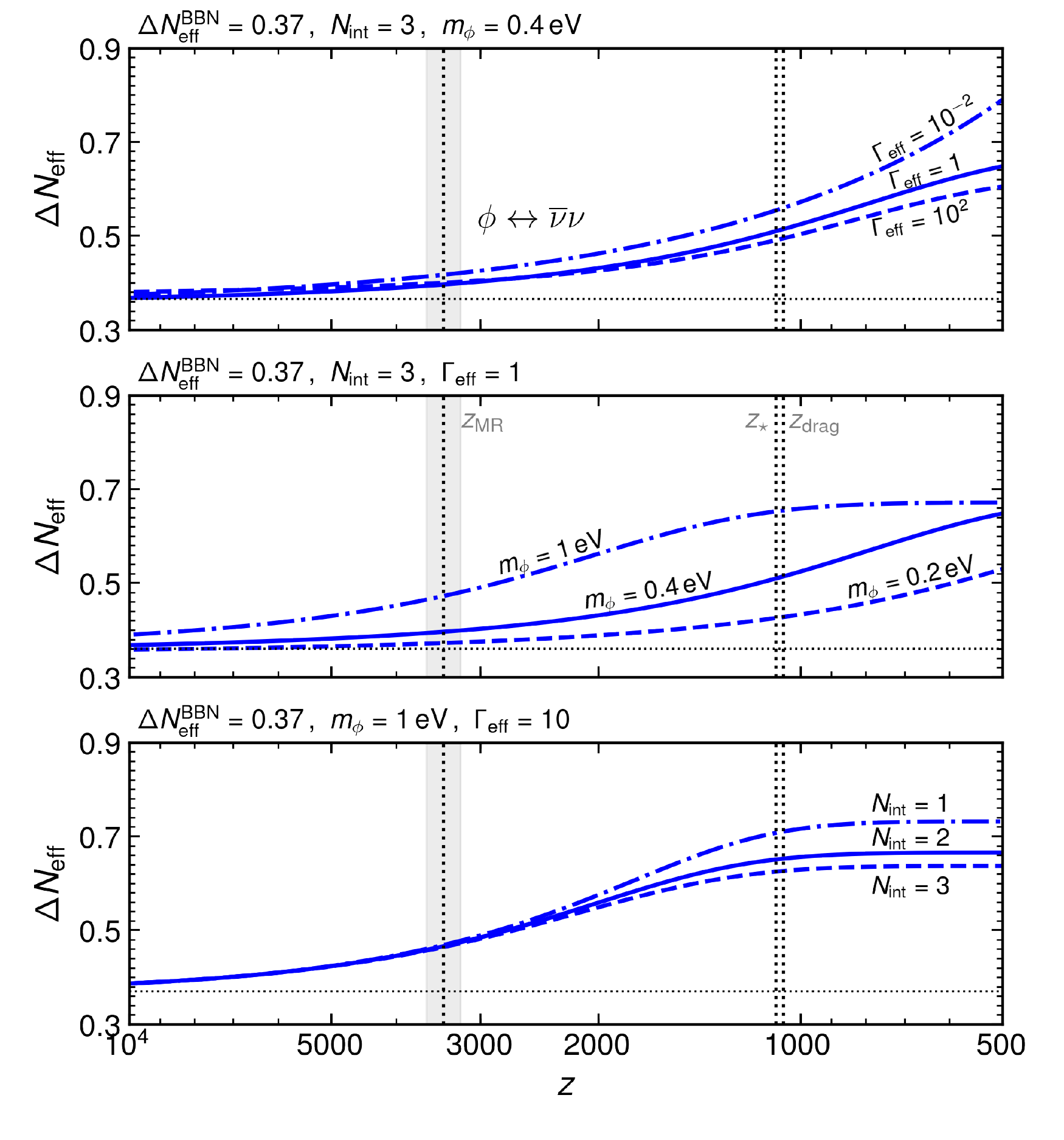}
	\vspace{-0.7cm}
	\caption{Evolution of $\Delta N_{\rm eff}$ as a function of redshift for the case $\Delta N_{\rm eff}^{\rm BBN} =0.37$. \textit{Upper panel:} evolution for a fixed value of $m_\phi$ for different values of the neutrino-majoron interaction rate $\Gamma_{\rm eff}$, see Eq.~\eqref{eq:gamma_eff2}. \textit{Middle panel:} evolution for a fixed $\Gamma_{\rm eff}$ and for three representative values of $m_\phi$. \textit{Lower panel:} evolution fixing $m_\phi$ and $\Gamma_{\rm eff}$ but with varying number of interacting neutrino species. We can appreciate that between matter-radiation equality and recombination $\Delta N_{\rm eff}$ can easily grow up to $\sim 0.6$ as a result of the $\phi \leftrightarrow \bar{\nu}\nu$ interactions.} \label{fig:DNeff_CMB}
\end{figure}

Given an initial primordial majoron population characterized by $\Delta N_{\rm eff}^{\rm BBN}$, we can solve for the evolution for the neutrino-majoron system by accounting for majoron decays and inverse neutrino decays $\phi \leftrightarrow \bar{\nu}\nu$. Notice that this system is controlled by three parameters: the number of interacting neutrinos $N_{\rm int}$, the effective decay width $\Gamma_{\rm eff}$ given in Eq.~\eqref{eq:gamma_eff2}, and the majoron mass. We highlight in Fig.~\ref{fig:DNeff_CMB} the impact of late-time neutrino-majoron thermalization on the expansion rate of the Universe (via the contribution to $\Delta N_{\rm eff}$), for various choices of these parameters. From the upper panel of Fig.~\ref{fig:DNeff_CMB} we can see that \textit{i)} the presence of neutrino-majoron interactions enhances $\Delta N_{\rm eff}^{\rm CMB}$ with respect to $\Delta N_{\rm eff}^{\rm BBN}$, and that \textit{ii)} for $\Gamma_{\rm eff} < 1$ the growth of $\Delta N_{\rm eff}$ is substantially larger (as mentioned above, this happens because for such values of $\Gamma_{\rm eff}$ majorons decay out of equilibrium at $T_\nu < m_\phi/3$).  Finally, from the lower panel of Fig.~\ref{fig:DNeff_CMB} we can see the impact of accounting for the three possible values of $N_{\rm int}$. The results are fairly similar, but perhaps contrary to what could be expected,  $\Delta N_{\rm eff}$  is larger for smaller $N_{\rm int}$. This happens simply as a result of equilibrium thermodynamics: smaller values of $N_{\rm int}$ lead to a more degenerate final state $\nu$ population  (because in practice more neutrinos have gone into that sector), and therefore have a larger energy density.

An enhancement of the expansion rate prior and close to recombination has long been appreciated as a key ingredient in models attempting to resolve the $H_0$ tension. The reason being that the characteristic angular size of fluctuations in the CMB has been measured with very high precision, $\sim 0.03\%$~\cite{planck}. For a flat FLRW Universe, this angular scale can be written as
\begin{align}
\theta_s = \left[\int_{z_{\star}}^{\infty} \frac{c_s(z)}{H(z)}dz \right]\bigg/\left[ \int_0^{ z_{\star}}\frac{dz}{H(z)}\right]\,,
\end{align}
where $z_{\rm \star}$ represents the redshift of last scattering, $c_s(z)$ is the photon-baryon sound speed, and $H(z)$ is the expansion rate of the Universe. Given that $\theta_s$ is measured to such high precision, and that $c_s(z)$  and $z_{\rm \star}$ are constrained via alternative observables, the most clear modification that allows for larger values of $H_0$ is to enhance $H(z)$ prior to recombination~\cite{Knox:2019rjx}. Using the approximate relationship between $\Delta N_{\rm eff}$ at recombination and $H_0$ derived in
\cite{Vagnozzi:2019ezj}:
\begin{align}
H_0  \simeq \left(67.5+ 6.2\,\Delta N_{\rm eff}^{\rm CMB}\right)\,\text{km}/\text{s}/\text{Mpc}\, ,
\end{align}
one can see that fully resolving the $H_0$ tension requires values of $\Delta N_{\rm eff}^{\rm CMB} \sim 1$. 
The problem is that values of $\Delta N_{\rm eff}^{\rm CMB} \gtrsim 0.3$ are disfavored by Planck data, regardless of whether this radiation is dark and free-streaming~\cite{planck} or strongly interacting~\cite{Blinov:2020hmc}. Our set up, however, is completely different from either of these cases because the majoron-neutrino interactions are only efficient for a limited period of time, roughly between $3\, m_\phi \gtrsim T_\nu \gtrsim m_\phi/10$.  For the mass range of interest, this implies majoron-neutrino interactions will only alter CMB multipoles $\ell \lesssim 1000$. This effect allows for an additional increase in $\Delta N_{\rm eff}$ relative to that of $\Lambda$CDM without spoiling the fit to the Planck observations.

\vspace{0.15cm}
\begin{center} \textbf{\small B. Planck Analysis}  \end{center}
\vspace{0.1cm}

To study in detail the effect of majoron-neutrino perturbations on the CMB we have modified the cosmological Boltzmann code {\tt CLASS}~\cite{Blas:2011rf,Lesgourgues:2011re}. In order to analyze the subsequent evolution of majorons as relevant for CMB observations we shall make a number of approximations, which we enumerate in what follows for the sake of clarity. First, we shall neglect neutrino masses in the evolution of the background energy density and at the level of the cosmological perturbations. We take this approximation because including neutrino masses at the perturbation level is rather complicated, see~\cite{Barenboim:2020vrr}. Nevertheless, this approximation is well justified provided that neutrino masses are at the level $\sum m_\nu \lesssim 0.2\,\text{eV}$. This being said, large neutrino masses may be capable of reducing the value of $\sigma_8$, as was the case for the strongly interacting neutrino solution (see~\cite{Kreisch:2019yzn}). Second, for the purpose of describing the neutrino-majoron perturbations we assume that they form a single coupled fluid. This is a good approximation because all species are effectively relativistic except when the majorons decay. Their contribution to the equation of state of the system is always below $13\%$ for the models presented, and typically $\lesssim 5\%$ for the parameter space of interest\footnote{This effect is larger for $N_{\rm int} = 1$, which is why we choose not to display the results for this model. We note, however, that adopting this same assumption for the $N_{\rm int} = 1$ model yields posteriors that are nearly identical to the $N_{\rm int} = 2$ and 3 models.}. Finally, we approximate the collision term at the perturbation level by the relaxation time approximation~\cite{Hannestad:2000gt}. This approach has been shown to be accurate for scenarios in which the interacting particles subtend large angles after collisions in the plasma~\cite{Oldengott:2017fhy}. In the parameter space we consider, majorons will only interact once they are mildly non-relativistic, by which time the typical angular separation of neutrinos in the cosmic frame is large, $\theta > 10^{\circ}$, which justifies our approach. Although it is beyond the scope of this paper to account for the exact collision term, we refer to~\cite{Barenboim:2020vrr} for a recent study dealing with the exact collision term in the context of invisible neutrino decays. 

\begin{figure}
	\includegraphics[width=0.48\textwidth]{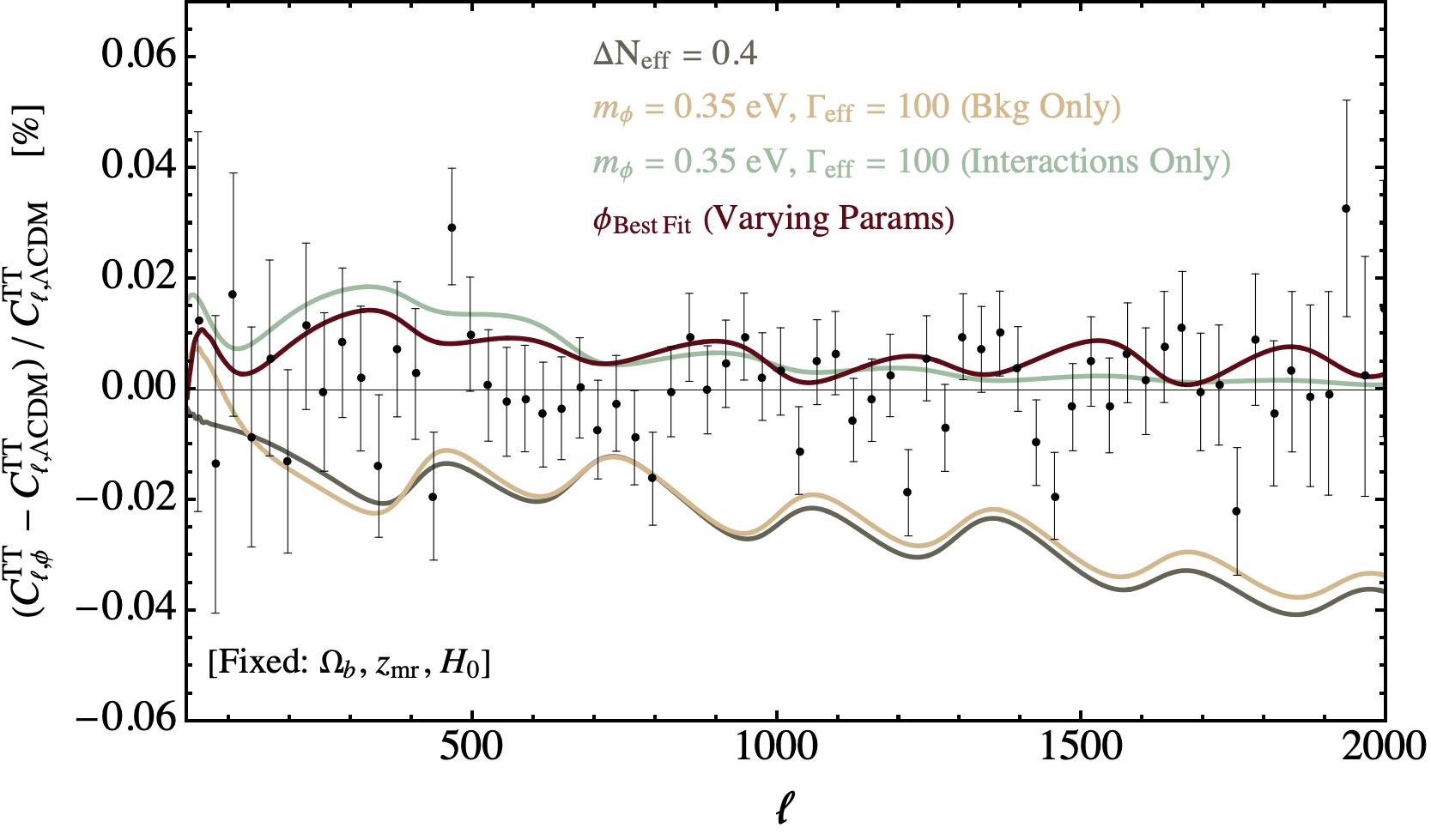}
	\caption{Relative difference between TT power spectrum compared in various models and that of $\Lambda$CDM, shown in comparison with Planck 2018 residuals. Models shown include $(i)$ constant $\Delta N_{\rm eff} = 0.4$ (grey), $(ii)$ primordial majoron with $T_d = 50$ MeV ($\Delta N_{\rm eff}^{\rm BBN} \equiv 0.37$), $N_{\rm int} = 2$, $m_\phi = 0.35$ eV and $\Gamma_{\rm eff} = 10^2$, including only the background (yellow), only the damping of free streaming (green), and the best-fit point (red). With the exception of $\phi_{\rm Best Fit}$, all models use $\Lambda$CDM best-fit parameters, adjusted to ensure $\Omega_b$, $z_{mr}$, and $H_0$ are constant. Notice that the best fit majoron cosmology has $\chi^2_\phi - \chi^2_{\Lambda \text{CDM}} = -4.5$, implying a better fit than $\Lambda$CDM to Planck+BAO data.}\label{fig:CLs}
\end{figure}

\vspace{0.4cm}
\newpage 

In Fig.~\ref{fig:CLs} we show how the various effects described here effect the TT power spectrum. Specifically, we isolate each effect, and show the residuals relative to that of $\Lambda$CDM for a majoron with $m_\phi = 0.35$ eV and $\Gamma_{\rm eff} = 10^2$. These lines are produced by fixing $\Omega_b$, $z_{mr}$, and $H_0$ to their  $\Lambda$CDM values. We compare the background only contribution to a model in which $\Delta N_{\rm eff}$ is fixed to $0.4$ to illustrate that the late-time thermalization only effects low multipoles. We also show the best-fit majoron cosmology for $N_{\rm int} = 2$ and $T_d = 50$ MeV in red ($\Delta N_{\rm eff}^{\rm BBN} = 0.37$), which illustrates the high quality fit obtained from the MCMCs. From Fig.~\ref{fig:CLs} we can appreciate that the neutrino-majoron interactions act as to partially cancel the background contribution at low multipoles.

To analyze in detail the cosmological implications of our scenario, we perform a MCMC using {\tt MontePython}~\cite{Audren:2012wb,Brinckmann:2018cvx} including {\emph{only}} {\tt Planck2018+BAO} data~\cite{planck,Aghanim:2019ame}\footnote{{\tt Planck2018} data~\cite{planck} includes the high-$\ell$ and low-$\ell$ (temperature and polarization) and lensing likelihoods~\cite{Aghanim:2019ame}. In this work we use BAO data that includes the 6DF galaxy
survey~\cite{Beutler:2011hx}, the MGS galaxy sample of SDSS~\cite{Ross:2014qpa}, and the CMASS and LOWZ galaxy samples of BOSS DR12~\cite{Alam:2016hwk,Vargas-Magana:2016imr,Ross:2016gvb,Beutler:2016ixs}, as used in the fiducial Planck analysis.}\footnote{We have also run an analysis including Pantheon type Ia supernova data, and have found that the preferred value of $H_0$ is surprisingly slightly shifted to yet higher values, and the $\Delta \chi^2$ with respect to LCDM is reduced. This emphasizes the robust nature of the obtained fit. }. For each of our analyses we vary the standard cosmological parameters and nuisance parameters in the same way as the Planck collaboration in their legacy analysis~\cite{planck}. For the majoron mass and interaction we use log-scale priors with the following ranges\footnote{We do not consider values $\Gamma_{\rm eff} < 10^{-2}$ because for such small interaction strengths majoron-neutrino interactions cannot significantly alter neutrino free-streaming. In addition, it was found that for models with a decoupling temperature $T_d = 30$ MeV, larger priors on $\Gamma_{\rm eff}$ were needed, and thus in these cases we adopt a log flat prior over the range $[10^{-2}, 10^{4}]$. }: 
\begin{align}
m_\phi  &= (0.1-10^3) \,\text{eV}\,,\\
\Gamma_{\rm eff}  &= 10^{-2}-10^2 \, .
\end{align}
We note that the lower limit on $m_\phi$ is imposed on the physicality of our model. It corresponds to the minimal mass for which the majoron can decay into the most massive neutrino $m_\phi \geq 2 \sqrt{|\Delta m_{\rm atm}^2|} = 0.1\,\text{eV}$. For smaller masses the majoron could potentially participate in the process of neutrino decay leading to a very different phenomenology~\cite{Barenboim:2020vrr,Escudero:2020ped,Chacko:2019nej,Chacko:2020hmh,Escudero:2019gfk,Hannestad:2005ex}. This, however, will in fact not occur within the framework of the singlet majoron model considered here, see~\cite{Schechter:1981cv}. 

In our runs we do not vary the initial primordial majoron population, i.e. $\Delta N_{\rm eff}^{\rm BBN}$ or equivalently $T_d$, but instead we explore some representative values expected to arise from low-scale leptogenesis; this is because of the difficulty in computing the thermodynamic evolution on-the-fly, something the authors hope to improve upon in future work.  We also explore  scenarios with $N_{\rm int} = 1\,,2\,,3$ interacting neutrino species. Nevertheless, in what follows, we concentrate on $N_{\rm int} = 2$ for concreteness because our runs show that the posteriors are fairly independent of  whether $N_{\rm int} = 1\,,2\,,3$.

In Fig.~\ref{fig:posterior_text} we show the resulting posterior (see Appendix~\ref{app:majorons_CMB} for the full results) for analyses with $N_{\rm int} = 2$ and $\Delta N_{\rm eff}^{\rm BBN} = 0.37$ in green, and $\Lambda$CDM in grey. For comparison, we also plot the one and two sigma posterior from local measurements of $H_0$ performed by the SH$_0$ES collaboration, which find $H_0 = 73.2 \pm 1.3$ km/s/Mpc~\cite{Riess:2020fzl}.

From our analysis of the fiducial model we obtain a number of relevant results:
\begin{enumerate}
\item Majoron cosmologies with a primordial population at the level of $\Delta N_{\rm eff}^{\rm BBN} \simeq 0.4$ lead to substantially larger values of $H_0$ than $\Lambda$CDM:
\begin{align}
H_0 = 70.2\pm0.6 \,\text{km/s/Mpc} \, .
\end{align}
\item With respect to $\Lambda$CDM, there are $\sim 1\sigma$ upward shifts on other cosmological parameters such as $n_s$, $\Omega_b h^2$, and $\Omega_{\rm cdm} h^2$.  Importantly, the observed shift in the preferred value of $\Omega_b h^2$ is in agreement with observations from BBN (see \App{app:bbn} for additional details). 
\item The majoron mass needed to obtain these rather large values of $H_0$ is bounded from above to be: 
\begin{align}\label{eq:mphi_range}
m_\phi \lesssim \, 1.3 \, {\rm eV} \, \hspace{.1cm} {\rm at} \hspace{.1cm} 2 \sigma\, \, ,
\end{align} 
and is bounded from below at $m_\phi > 2\sqrt{|\Delta m_{\rm atm}^2|} = 0.1\,\text{eV}$. We note that these masses are in agreement with the expectations from Fig.~\ref{fig:DNeff_CMB}. In addition, we note that this upper limit highlights that primordial populations of majorons with larger than eV-scale masses that decay well before recombination are not favored by Planck legacy data. 
\item The preferred region of parameter space for majoron interactions is
\begin{align}\label{eq:Gammaeff_range}
\Gamma_{\rm eff} \gtrsim 0.03 \, \hspace{.1cm} {\rm at} \hspace{.1cm} 2 \sigma\, \, .
\end{align} 
\item The Planck legacy data points to scales of spontaneous lepton number breaking: 
\begin{align}\label{eq:vL_range}
v_L  \lesssim 1\,\text{TeV} \,,
\end{align} 
where this number is obtained by taking Eq.~\eqref{eq:lambda_def} with $m_\nu = 0.05\,\text{eV}$, but we note that the bound on $v_L$ could be as large as $2-3$ TeV for more massive neutrinos.

\item The best-fit point for of the fiducial model has a $\chi^2$ of:
\begin{align}
\chi^2 |_{\Delta N_{\rm eff}^{\rm BBN} = 0.37}^{\phi} - \chi^2|_{\Lambda {\rm CDM}}&= -4.5 \, .
\end{align} 
and corresponds to
\begin{align}\label{eq:bfpoints}
m_\phi = 0.35\,\text{eV}\,,\,\,\, \Gamma_{\rm eff} = 67.6\, , \,\,\, v_L \simeq 330\,\text{GeV}  \,. \!\!\! \!\!\! \!\!\!
\end{align}  
These $\Delta \chi^2$ values highlight that for moderately large $\Delta N_{\rm eff}^{\rm BBN}$ the fit to Planck+BAO data can be improved relative to $\Lambda$CDM. However, our other analyses show that the fit becomes degraded for primordial populations with $\Delta N_{\rm eff}^{\rm BBN} \gtrsim 0.5$. Notice that these values of $\Delta N_{\rm eff}$ are excluded at more than 2$\sigma$ if the damping of neutrino free streaming is not included~\cite{planck}.
\end{enumerate}

\begin{figure}
	\includegraphics[width=0.495\textwidth]{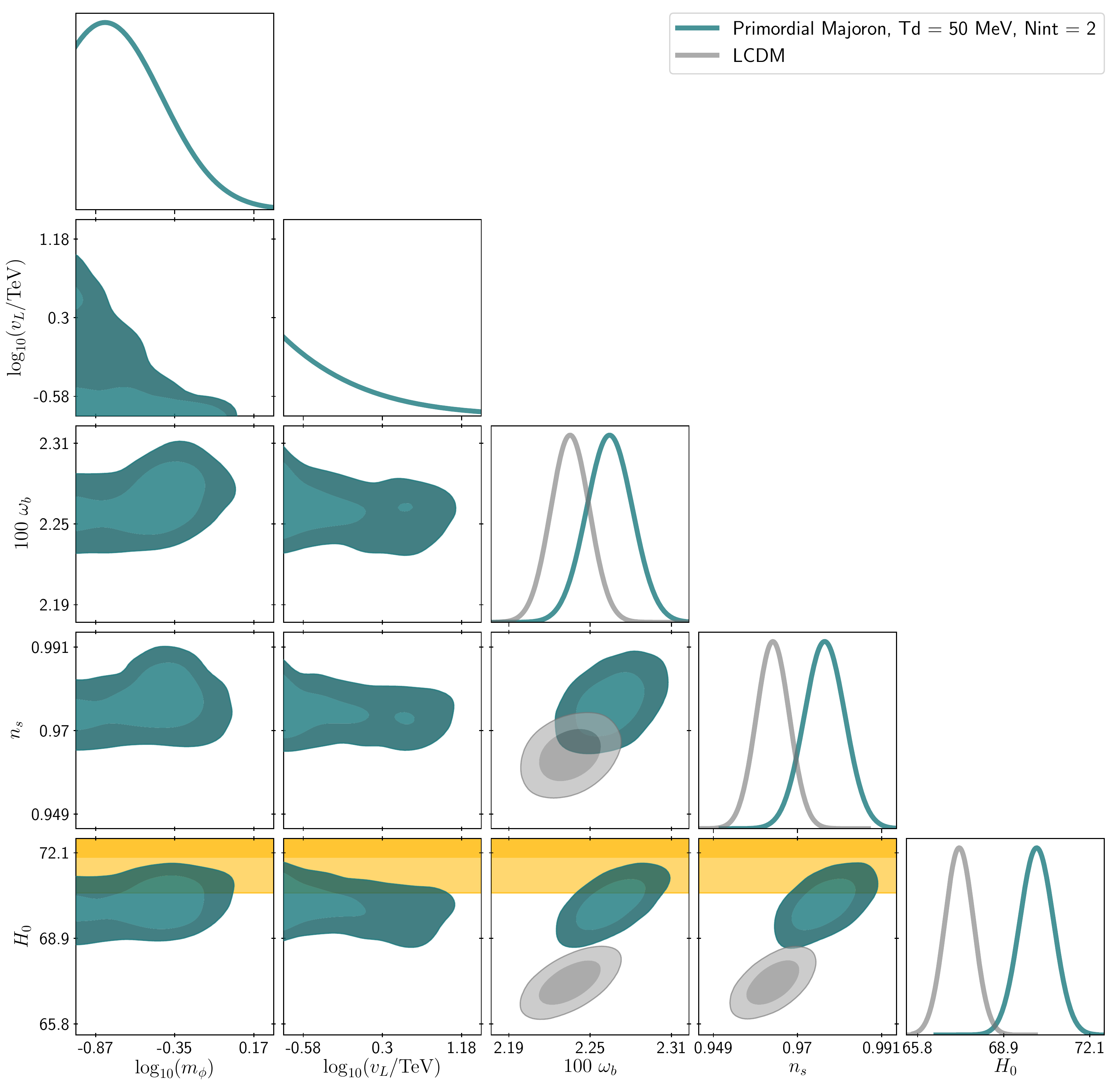}
	\caption{MCMC fit to Planck2018+BAO data for a majoron cosmology with $N_{\rm int} = 2$ (green) with $T_d = 50$ MeV, and for $\Lambda$CDM (grey). One and two sigma posteriors on $H_0$ from the SH$_0$ES collaboration are presented in yellow. }\label{fig:posterior_text} 
\end{figure}

\begin{figure}
	\includegraphics[width=0.48\textwidth]{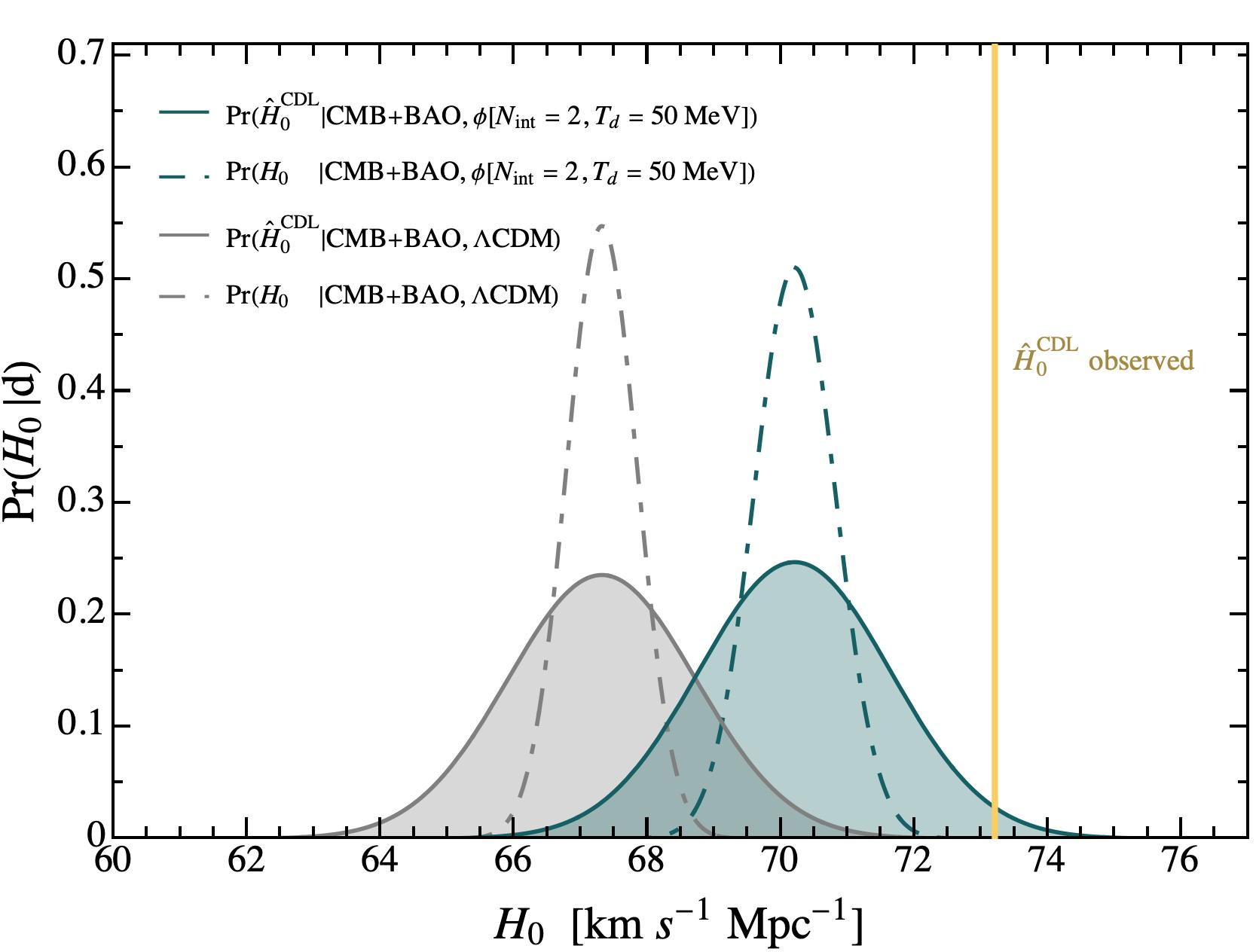}
\vspace{-0.45cm}
\caption{\label{fig:ppd} Probability Predictive Distributions (PPD) (shaded) for the SH$_0$ES measurement of $H_0$, conditioned on Planck2018+BAO data and assuming either  $\Lambda$CDM (grey) or the leptogenesis-motivated majoron model with $N_{\rm int} = 2$ and a decoupling temperature of 50 MeV (green).
The posterior on $H_0$ for each model is shown for comparison with dot-dashed lines, and the central value of the SH$_0$ES measurement in shown for comparison in yellow.}
\end{figure}

\vspace{0.15cm}
\begin{center}\textbf{\small C. Implication for the Hubble Tension}  \end{center}
\vspace{0.1cm}

In order to determine the significance with which the leptogenesis motivated majoron resolves the Hubble tension, we use the so-called posterior predictive distribution~\cite{gelman2013bayesian} (PPD) as proposed in \cite{Feeney:2018mkj}. Simply put, the idea is to determine the distribution of expected measurements obtained from new data $(\bf d^\prime)$ given a set of observed data $(\bf d)$ and a model $(I)$; in the context of the Hubble tension, this is equivalent to asking the question: given the value of $H_0$ observed by CMB and BAO data within a particular cosmology (taken here to be either $\Lambda$CDM or the primordial majoron cosmology), how likely is it for the SH$_0$ES collaboration to measure some other value $\hat{H_0}^{\rm CDL, \, obs}$? The PPD is obtained by averaging the likelihood of the new data over the posterior of the parameters $(\bf \theta)$ of the existing data, \ie
\begin{equation}
	Pr({\bf d^\prime}| {\bf d}, \, I) = \int d{\bf \theta} \, Pr({\bf{d^\prime}} | {\bf{\theta}}, I) \, Pr(\theta | {\bf{d}}, I) \, . 
\end{equation} 
Importantly, there is no assumption that the two datasets under consideration are consistent. Rather, the PPD allows one to determine whether the second dataset is a likely draw from the first. We apply this method using the measurement of $H_0$ by the SH$_0$ES collaboration, for which we adopt a Gaussian likelihood with central value $73.2$ km/s/Mpc and uncertainty $\sigma = 1.3$ km/s/Mpc~\cite{Riess:2020fzl}. We plot the PPD for $\Lambda$CDM and the leptogenesis-inspired majoron model with $N_{\rm int} = 2$ in \Fig{fig:ppd} (shaded regions). The central value of the SH$_0$ES measurement is also shown for comparison (vertical yellow line).

The potential tension between the CMB+BAO measurement and that of SH$_0$ES can then be obtained using the PPD ratio, defined as
\begin{equation}
	\rho_{{\rm PPD}} = \frac{Pr(\hat{H}_0^{\rm CDL, \, obs} | {\bf d}, I)}{{\rm max}\left[Pr(\hat{H}_0^{\rm CDL} | {\bf d}, I) \right]} \, ,
\end{equation}
where $\hat{H}_0^{\rm CDL, \, obs}$ is the value of the Hubble constant observed by SH$_0$ES by using the Cepheid Distance Ladder (CDL). The PPD ratio can be interpreted as a lower bound on posterior probability of the hypothesis that the CMB+BAO data and that of SH$_0$ES arise from the same value of $H_0$ without unaccounted for systematics. In the case of $\Lambda$CDM, we find a value of $\rho_{\rm PPD} \sim 10^{-3}$, implying the probability that the SH$_0$ES measurement is unaffected by systematics and appears only from statistical fluctuations, is roughly $\sim 0.1 \%$. In the majoron model, on the other hand, we find a value of $\rho_{\rm PPD} \sim 10\%$.  Admittedly, this metric does not account for the fact that additional model parameters have been introduced; nevertheless, the statistical tension between these datasets in the context of $\Lambda$CDM is sufficiently large that novel physics must now be considered, and thus it is perhaps more appropriate to compare the number of  model parameters introduced only between proposed solutions to the Hubble tension. 

A potential criticism of this treatment is that the decoupling temperature is fixed, rather than scanned, and thus has preferentially selected a large value of $\Delta N_{\rm eff}$\footnote{Note that implementing a scan of $T_d$ is quite complicated and time consuming, as the out-of-equilibrium evolution of the majoron must be solved on the fly in {\tt CLASS}~\cite{Blas:2011rf,Lesgourgues:2011re}. }; the current treatment can be interpreted as adopting a strong prior on the ARS leptogenesis-motivated decoupling temperature. As shown in Appendix~\ref{app:majorons_CMB}, the $\Delta \chi^2$ of the best-fit is approximately equivalent to, and actually slightly below, that of $\Lambda$CDM. It is thus straightforward to understand that the effect of scanning $T_d$ would simply be to broaden the posterior on $H_0$ about a value of $\sim 70$ km/s/Mpc, extending at lower values to what is found in $\Lambda$CDM\footnote{The limit in which $T_d \rightarrow \inf$, $\Gamma \gtrsim 1$, and $m_\phi \gg \text{eV}$ is as close as this model can be to recovering $\Lambda$CDM, however even in this limit energy is injected at the level of $\Delta N_{\rm eff} \simeq 0.12$. } and potentially extending in the low decoupling temperature limit to a value  $H_0 \sim 73 $ km/s/Mpc (as shown in~\cite{Escudero:2019gvw}). Including a prior on $H_0$ from \eg the SH$_0$ES measurement, however, would shift the posterior to values of $H_0$ larger than that obtained here, and remove the part of the posterior near the $\Lambda$CDM value.

\section{Summary and Conclusions}\label{sec:con}

\vspace{0.35cm}

Majorons with $\sim \text{eV}$ masses can thermalize with neutrinos prior and close to recombination, damping neutrino free streaming, altering the energy density stored in the neutrinos themselves. These effects manifest in the CMB power spectrum in a manner that partially cancels, however there is a residual scale-dependent phase shift that leads to a preference for larger values of $H_0$, $n_s$, $\Omega_b h^2$ and $\Omega_{\rm cdm} h^2$. This motivated the authors' previous study~\cite{Escudero:2019gvw}, where it was shown that a majoron with a mass $m_\phi \sim \text{eV} $ arising from a the spontaneous breaking of a lepton number symmetry at scales $v_L \sim \mathcal{O}(1)$ TeV could reduce the Hubble tension to the $\sim 2\sigma$ level, however only if additional dark radiation was present at the level of $\Delta N_{\rm eff} \sim 0.5$.

\vspace{0.35cm}

In this work we have investigated the extent to which a primordial population of majorons arising from low-scale models of leptogenesis can source the additional radiation required to ameliorate the $H_0$ tension. We have found that so long as the Higgs' portal coupling is sufficiently small ($|\lambda_{\phi H}| \lesssim 10^{-7}$) so as to avoid thermalizing the new scalars at high temperatures and that the lepton number phase transition occurs at $T>10^{4}-10^6\,\text{GeV}$, ARS leptogenesis in the singlet majoron model will proceed as normal. Given the sterile neutrino masses required for ARS leptogensis and the range of interaction strengths required for majorons to thermalize at late times, sterile neutrinos will inevitably create a thermal majoron population at temperatures $T \gtrsim M_N$. This primordial majoron population will eventually decouple as sterile neutrinos decay $T\sim M_N/10$, naturally sourcing a contribution of  $\Delta N_{\rm eff} \sim 0.4$ (at BBN),  see Fig.~\ref{fig:neff_BBN}. Given that a large population of primordial majorons can be produced in the early Universe, the question becomes whether the late-time evolution of the majoron-neutrino system behaves similarly to the analysis of~\cite{Escudero:2019gvw} (since the time-dependence of the interaction rate has been modified), and whether the result is sensitive to the number of interacting neutrinos.  Our results indicate that all three scenarios (\ie that in which one, two, or all three neutrinos interact) favor similar parameter space, provide a similar shift in $H_0$, and yield comparable $\chi^2$ values to CMB and BAO data. The preferred value of $H_0$ of these models using only {\tt Planck2018+BAO} data is found to be $\sim 70.2 \pm 0.6$ km/s/Mpc, still below the value reported by the SH$_0$ES collaboration $73.2\pm 1.3$ km/s/Mpc~\cite{Riess:2020fzl}, but compatible at the $\sim 10\%$ level (to be compared with $\sim 0.1\%$  in $\Lambda$CDM). It remains to be seen whether a more robust analysis including non-zero neutrino masses, both in the evolution of the background and at the level of the perturbations, would alter the results in a significant manner.

\newpage
To summarize, we have shown that the Hubble tension could be a signal of low-scale leptogenesis and neutrino mass generation within a simple and well-motivated neutrino mass model. In this set up, lepton number is a global symmetry that is spontaneously broken at an energy scale $v_L$. This generates Majorana masses for sterile neutrinos, which in turn via the type-I seesaw lead to an understanding of the small neutrino masses. 
In addition, upon breaking of $U(1)_L$ a pseudo-Goldstone boson appears on the spectrum, the majoron $\phi$. This particle is naturally very light and $\text{eV}$ masses for it are motivated from Planck-scale suppressed operators that explicitly break lepton number, $m_\phi \lesssim v_L \sqrt{v_L/M_{\rm Pl}}$.  Furthermore, in this set up, the interactions between neutrinos and majorons are extremely feeble, $\lambda \sim m_\nu/v_L \sim 10^{-13}$. Intriguingly, these very small couplings and masses precisely correspond to $\phi \leftrightarrow \bar{\nu}\nu$ processes turning on right before recombination for $v_L \sim v_H$. Indeed, our Planck legacy data analysis shows that the preferred region of parameter space within a primordial majoron cosmology is:
\begin{align}
\Delta N_{\rm eff}^{\rm BBN} &\sim 0.4\,,\\
m_\phi &\sim   (0.1-0.8)\,\text{eV}\,, \\
v_L &\sim (0.03-2)\,\text{TeV} \,,
\end{align}
which point towards
\begin{align}
H_0 = 70.2\pm0.6\,\text{km/s/Mpc}\,.
\end{align}
Finally, given that the preferred scale of lepton number breaking is $\lesssim 2\,\text{TeV}$ this naturally points to GeV-scale sterile neutrinos. In particular, these sterile neutrinos can produce the right primordial majoron population (see Fig.~\ref{fig:neff_BBN}) and also generate via their oscillations the sufficient leptonic CP asymmetry in the early Universe. Therefore providing an understanding of the observed asymmetry between matter and antimatter.

\section{Outlook}\label{sec:outlook}
We have presented a connection between the Hubble tension, leptogenesis and the origin of neutrino masses. In what follows, we enumerate a list of aspects that fall beyond the scope of this study but that we believe deserve future attention:
\begin{enumerate}
\item \textit{Refined cosmological analysis}. As discussed in the main text, we have made three approximations which simplify the treatment of the majoron cosmology: \textit{i)} We have neglected the effect of neutrino masses in the evolution of both the background and the perturbations of the Universe. This is well-justified for $\sum m_\nu \lesssim 0.2\,\text{eV}$ as Planck CMB observations are not sensitive to such small neutrino masses. However, an actual analysis could potentially reveal the preference of higher neutrino masses which will yield a smaller value of $S_8$ as relevant for weak lensing measurements of this parameter, see e.g.~\cite{Jedamzik:2020zmd}. \textit{ii)} We have modeled the majoron-neutrino perturbations by making use of the relaxation time approximation for the collision term. As argued in Section~\ref{sec:cosmo} we believe that this should be a good approximation because in our parameter space of interest majorons interact with neutrinos once they become mildly non-relativistic, implying that the final state particles will be roughly isotropically produced (as assumed by the relaxation approximation to capture the effect on the anisotropic stress). It would, however, be very interesting to explore the phenomenology accounting for the exact collision term. \textit{iii)} We have treated the neutrino-majoron system as a single coupled fluid.  While this assumption holds in the relativistic limit, the isotropisation of the fluid decreases as majorons become non-relativistic and decay.  It is thus possible that relaxing this assumption, and properly incorporating the decay hierarchy, could potentially reduce the maximum allowed value of $\Delta N_{\rm eff}$. We hope to refine this treatment, and properly assess the importance of this effect, in future work.

\item \textit{ARS Leptogenesis in the singlet majoron model}. In this work (see also~\cite{Caputo:2018zky}) we have derived a minimal set of conditions necessary for the singlet majoron model not to spoil successful ARS leptogenesis. In order to robustly determine the parameter space of interest more rigorous calculations would be necessary\footnote{We note that baryogenesis/leptogenesis has been explored in the majoron model in the context of electroweak baryogenesis~\cite{Cohen:1990it,Cohen:1990py}, thermal leptogenesis~\cite{Sierra:2014sta}, and resonant leptogenesis~\cite{Pilaftsis:2008qt}. In addition, we note that the CP violating decays of Higgs doublets into sterile neutrinos in the early Universe can yield relevant lepton asymmetries that could (depending upon the mass degeneracy) dominate over the contribution arising from oscillations~\cite{Hambye:2016sby,Hambye:2017elz}.}. We note that given the temperature dependence of $M_N$ is this model, such analysis may yield even more favorable conditions for the production of a lepton asymmetry from sterile neutrino oscillations.

\item[3.] \textit{Collider Detectability}. Although in this study we have exhausted the cosmological implications of majorons and their companion neutrinos and sterile neutrinos, we have not discussed potential signals at laboratory experiments. From the collider perspective, it appears quite difficult to test the scalar sector of the theory given the smallness of the Higgs portal coupling and the smallness of the neutrino-majoron couplings. However, in the context of sterile neutrinos there are some possibilities. In our set up, sterile neutrinos decay invisibly into a neutrino and a majoron (see Eq.~\eqref{eq:ratio}). This means that typical searches at beam-dump experiments (see e.g.~\cite{Alekhin:2015byh}) will not have sensitivity to this model. The best avenue to detect these GeV-scale sterile neutrinos may be to look for $K/\pi\to \ell  \,N$~\cite{Aguilar-Arevalo:2017vlf,CortinaGil:2021gga} decays, where the $N$ particles appears in the form of missing energy. While at the moment current experiments are sensitive to active-sterile neutrino mixings $\sim (1-2)$ orders of magnitude larger than in the naive seesaw limit (see Eq.~\eqref{eq:seesawmixing}), ongoing and upcoming experiments looking for these decays~\cite{Beacham:2019nyx} may be sensitive to the minimum mixing expected from the seesaw for $m_N < m_{K/\pi}-m_{\mu,\,e}$.
\end{enumerate}

In this work we have shown that the Hubble tension can be largely ameliorated in a simple framework that explains both the origin of the active neutrino masses (via the seesaw mechanism) and the baryon asymmetry of the Universe (via the ARS leptogenesis mechanism). This proposal may be exhaustively tested by both future cosmological observations and by looking in terrestrial experiments for the presence of GeV-scale sterile neutrinos, which are necessary to source both the primordial majoron population and a primordial lepton asymmetry.

\begin{center}\textbf{Acknowledgments}  \end{center}

We are grateful to Iv\'{a}n Esteban, Pilar Hern\'{a}ndez, Kevin Kelly, Manuel Masip, Vivian Poulin, Nuria Rius and Jordi Salvad\'{o} for very useful comments and discussions. ME is supported by a Fellowship of the Alexander von Humboldt Foundation.  SJW acknowledges funding from the European Research Council (ERC) under the European Union's Horizon 2020 research and 
innovation programme (Grant agreement No. 864035 - UnDark).
\vspace{0.2cm}

\bibliography{biblio}

\onecolumngrid
\appendix

\setcounter{equation}{0}
\setcounter{figure}{0}
\setcounter{section}{0}
\setcounter{table}{0}
\makeatletter
\renewcommand{\theequation}{A\arabic{equation}}
\renewcommand{\thefigure}{A\arabic{figure}}
\renewcommand{\thetable}{A\arabic{table}}


\section{Low-Scale Leptogenesis Within the Singlet Majoron Model}\label{sec:ARS_app}

As discussed in the main text there are three key elements that are required to maintain the efficiency of the ARS leptogenesis mechanism within the singlet majoron model:
\begin{enumerate}
\item Sterile neutrinos cannot thermalize with the Standard Model plasma prior to $T  \sim T_{\rm EW}$. This would lead to a vanishing total lepton number asymmetry, and thus also to a vanishing baryon asymmetry. 
\item CP violating sterile neutrino oscillations must occur when $t_{\rm osc} \sim 1/H$. This is a non-trivial requirement in the singlet majoron model since the oscillation rate is related to sterile neutrino masses, which are only generated upon breaking of the $U(1)_L$ symmetry. This amounts to ensuring that the $U(1)_L$ symmetry is spontaneously broken at temperatures $T> T_{\rm lepto} \sim 10^5\,\text{GeV} $ (see Eq.~\eqref{eq:TLepto}).
\item The CP violating oscillations at $T\sim T_{\rm lepto}$ should be coherent so that a primordial CP asymmetry can be generated efficiently. 
\end{enumerate}

The former requirement can be obtained by requiring that neither $\rho$ nor $\phi$ thermalize for $T > T_{\rm EW}$. The dominant production mechanism for these states arises via the Higgs-portal coupling  $\lambda_{\Phi H}$ in Eq.~\eqref{eq:V_phi}. At $T > T_{EW}$ the rate at which $\rho$'s can be produced via this interaction is given by  $\left<\Gamma\right> =  \lambda_{\Phi H}^2 T /(132\pi^3) $. We can express the number density of $\rho$'s produced relative to the thermal equilibrium value as
\begin{align}\label{eq:Xi_rho}
\xi_{\rho,\phi} (T) \equiv \frac{n_{\rho,\phi}}{n^{\rm eq}_{\rho,\phi}} = 3\times 10^{-7} \left[\frac{|\lambda_{\phi H}|}{10^{-7}}\right]^2 \left[\frac{10^6\,\text{GeV}}{T}\right]\,,
\end{align}
where we have defined  $\xi_{\rho,\phi} \equiv n_{\rho,\phi}(T)/n_{\rm MB}(T)$, with $n_{\rm MB}(T)= T^3/\pi^2$. Here, we have used the fact that $\frac{d\xi}{dT} = -\frac{\xi}{T} \frac{\left<\Gamma\right>}{H}$.  Requiring that $\phi$ and $\rho$ do not thermalize with the SM plasma at $T\sim 100\,\text{GeV}$ is  equivalent to requiring $\xi_{\rho,\phi}(T = 100 {\rm GeV}) < 1 $, which yields the following condition  on the Higgs portal coupling:
\begin{align}
|\lambda_{\Phi H}| < 5\times 10^{-7}\,\qquad \text{(no\,}\rho/\phi\,\text{thermalziation at}\, T\gtrsim100\,\text{GeV}) \,.
\end{align}

In order to assess the viability of the second requirement we must look into the dynamics of the $U(1)_L$ phase transition. This requires knowledge, however, on the pressure induced from the $N$, $\rho$ and $\phi$ states, which intrinsically depend on  $h_{\alpha i}$ and $\lambda_{\Phi H}$. Since we are assuming all species are produced from the plasma with $E\sim T$ and have very small abundances, we can express their distribution functions as  $f^{\rm F/B} = \xi(T) f^{\rm FD/BE}(T) $ where again $\xi(T) \equiv n_i / n_i^{\rm eq}$ (note that similar assumptions are made in conventional studies of ARS leptogenesis~\cite{Eijima:2018qke}). This choice is justified on the basis that the sterile neutrinos produced from the SM plasma have typical momentum $p \sim (0.5-3)\,T_\gamma $~\cite{Besak:2012qm}. Given our ansatz for the distribution functions, we can calculate the relevant sterile neutrino abundance $\xi_N(T)$ as a function of temperature,
\begin{align}\label{eq:Xi_N}
\xi_N (T)\equiv \frac{n_{N}}{n^{\rm eq}_N}   =1.3\times 10^{-6} \left[\frac{|h|}{4\times 10^{-8}}\right]^2 \left[\frac{10^6\,\text{GeV}}{T}\right]\,,
\end{align}
where once again we have made use of  the fact that $\frac{d\xi}{dT} = -\frac{\xi}{T} \frac{\left<\Gamma\right>}{H}$, where in this case $\Gamma \simeq 5\times 10^{-3}\,|h|^2\,T$\footnote{This rate is calculated using $\frac{\delta n_{N_i}}{\delta t} \sim 7\!  \times \!10^{-4} \,  \sum_\alpha |h_{\alpha i}|^2  \, T^4$~\cite{Besak:2012qm}, and taking the typical neutrino energy to be  $E\sim 2T$; collectively this implies $\frac{\delta \rho_{N_i}}{\delta t}\simeq 2T \frac{\delta n_{N_i}}{\delta t} $. The thermally averaged rate is then obtained via $\left<\Gamma\right> = \frac{1}{\rho_N} \frac{\delta \rho}{\delta t}$.} 
where $|h|^2 \equiv \sum_\alpha |h_{\alpha i}|^2 $, and we have normalized it to the seesaw limit case with $M_N = 1\,\text{GeV}$, see Eq.~\eqref{eq:seesaw_DiracY}.  

It is important to note that Eqs.~\eqref{eq:Xi_rho} and~\eqref{eq:Xi_N} do not necessarily give the true evolution of the number densities of $N$, $\rho$ and $\phi$ at very high temperatures. The reason being that we have not accounted for the possibility that processes such as $ \rho \leftrightarrow \phi \phi $, $\bar{N} N \leftrightarrow \rho\rho$, etc.  change the relative number densities of each of these species. Nevertheless, the total number densities of all of these states will remain unchanged, and thus  Eqs.~\eqref{eq:Xi_rho} and~\eqref{eq:Xi_N} represent upper bounds on the number densities. If interactions between these particles are efficient, all species will have roughly equivalent abundances, with the value give by the maximum of Eq.~\eqref{eq:Xi_rho} and  Eq.~\eqref{eq:Xi_N}.  

\begin{figure}
\centering
\begin{tabular}{ccc}
\includegraphics[width=0.31\textwidth]{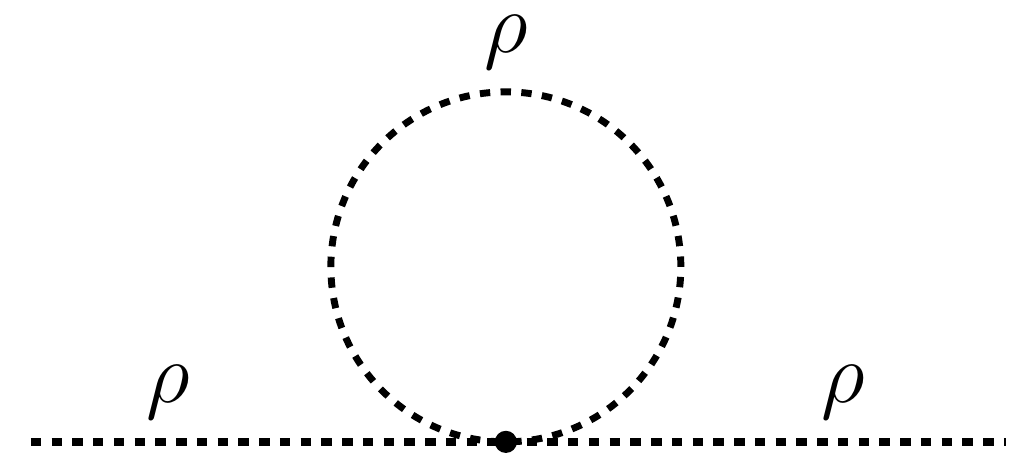} & \includegraphics[width=0.31\textwidth]{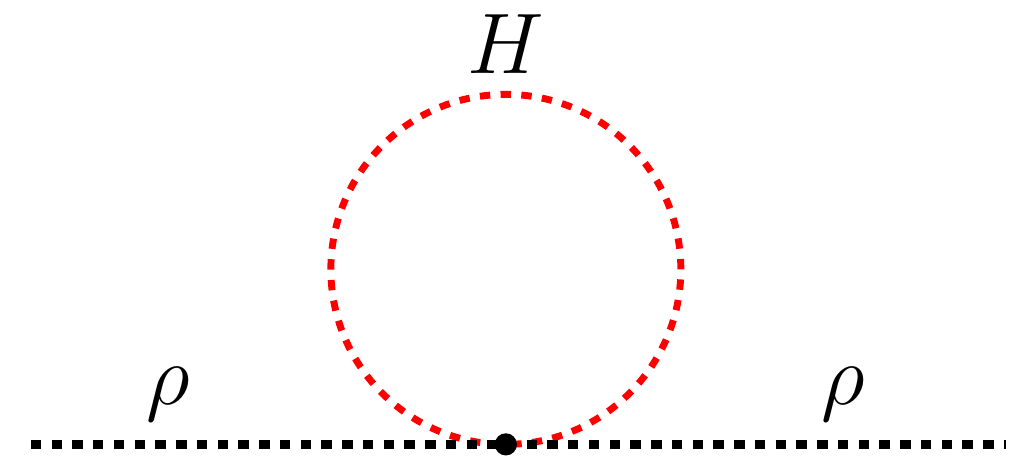} & \includegraphics[width=0.31\textwidth]{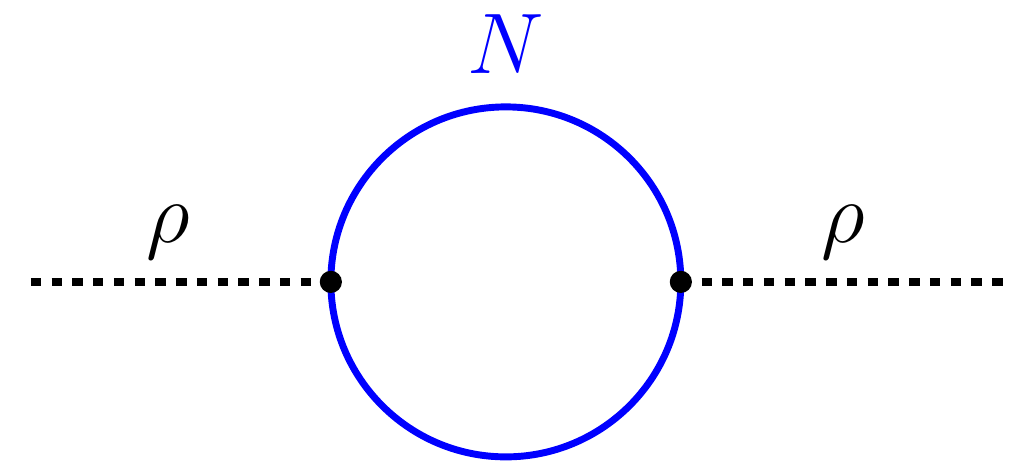} 
\end{tabular}
	\caption{Three relevant contributions to the thermal mass of the CP-even scalar $\rho$. See Eq.~\eqref{eq:m_phieff_each} for the actual contribution. }\label{fig:mrho_Thermal} 
\end{figure}

Since we know how $\xi_N$ and $\xi_\rho$ scale as a function of temperature, we can now study the evolution of the $U(1)_L$ phase transition. The 1-loop effective potential for the $U(1)_L$ sector is given by~\cite{Gorbunov:2011zz}
\begin{align}\label{eq:V_finite_T}
 V=& m_\rho^2|_{\rm eff}(T) \frac{1}{2} {\rho'}^2 + \frac{\lambda_\phi}{4} {\rho'}^4 - \frac{\lambda_{\Phi H}}{4} h^2\, {\rho'}^2 \,,
\end{align}
where we have defined $\rho' \equiv \rho + v_L$. In writing this expression we have neglected logarithmic contributions and terms linear in $T$, which are inherently small for the parameter space of interest -- as such, one can see that the $U(1)_L$ phase transition is of 2nd order. Here $m_\rho^2|_{\rm eff}(T) $ is the effective thermal mass of the $\rho$ scalar which receives contributions from the three self-energy diagrams in Fig.~\ref{fig:mrho_Thermal}. The contribution of each to the effective mass is
\begin{subequations}\label{eq:m_phieff_each}
\begin{align}
m_\rho^2|_{\rm eff}^\rho  &=T^2 \frac{1}{4} \, \lambda_\Phi \,\xi_\rho(T) \,, \\
m_\rho^2|_{\rm eff}^H  &= -T^2 \frac{1}{24} \, \lambda_{\Phi H} \,, \\
m_\rho^2|_{\rm eff}^N  &=  T^2\,\frac{1}{12}  \, \sum_i  \lambda_{N_i}^2 \xi_{N_i} (T) \, .
\end{align}
\end{subequations}
Summing over all the contributions and including the tree-level value we obtain
\begin{align}\label{eq:m_phieff}
m_\rho^2|_{\rm eff}(T)  &= -\lambda_\Phi v_L^2 + \frac{T^2}{12} \left[ 3 \lambda_\Phi \xi_\rho(T)   -  \lambda_{\Phi H}/2   +  \sum_i  \lambda_{N_i}^2 \xi_{N_i} (T) \right] \,.
\end{align}

In order to ensure sterile neutrinos oscillations proceed as normal we must ensure  $M_N\neq 0$ at $T_{\rm lepto} \sim 10^{4}-10^5 \,\text{GeV}$. This amounts to requiring the phase transition to occur at $T_c > T_{\rm lepto}$, where $T_c$ defines the temperature at which thermal corrections are subdominant to the tree-level value (such that the symmetry is broken by the vacuum). For each term in the thermal potential, we then derive constraints ensuring the aforementioned condition is met; this process gives
\begin{subequations}\label{eq:m_phieff_each_conditions}
\begin{align}
m_\rho^2|_{\rm eff}^\rho     \quad \longrightarrow\quad &|\lambda_{\Phi H}| <  4.6 \times 10^{-7}\, \frac{v_L}{1\,\text{TeV}} \sqrt{\frac{10^5\,\text{GeV}}{T_c}}\,,\quad (\xi_\rho = \xi_\rho) \,, \label{eq:m_Rho_R}  \\
m_\rho^2|_{\rm eff}^\rho     \quad \longrightarrow\quad &\sqrt{\sum_\alpha |h_{\alpha i}|^2} <  2 \times 10^{-7}\, \frac{v_L}{1\,\text{TeV}} \sqrt{\frac{10^5\,\text{GeV}}{T_c}}\,,\quad (\xi_\rho = \xi_N) \,,  \label{eq:m_Rho_N}   \\
m_\rho^2|_{\rm eff}^H     \quad \longrightarrow\quad &|\lambda_{\Phi H}| <  10^{-5} \left[\frac{m_\rho}{100\,\text{GeV}}\right]^2  \, \left[\frac{10^5\,\text{GeV}}{T_c}\right]^2\, ,  \label{eq:m_H}  \\
m_\rho^2|_{\rm eff}^N     \quad \longrightarrow\quad &\lambda_{N_i} <  0.7\, \frac{m_\rho}{100\,\text{GeV}} \frac{4\times 10^{-8}}{\sqrt{\sum_\alpha |h_{\alpha i}|^2}} \sqrt{\frac{10^5\,\text{GeV}}{T_c}}\,,\quad (\xi_N = \xi_N) \,,   \label{eq:m_N_N}  \\
m_\rho^2|_{\rm eff}^N     \quad \longrightarrow\quad &\lambda_{N_i} <  0.6\, \frac{m_\rho}{100\,\text{GeV}} \frac{10^{-7}}{|\lambda_{\phi H}|} \sqrt{\frac{10^5\,\text{GeV}}{T_c}}\,,\quad (\xi_N = \xi_\rho)\,. \label{eq:m_N_R} 
\end{align}
\end{subequations}
We have differentiated here between the case in which the number density of $N's$ is equal to that expected from the Dirac Yukawas and from the Higgs portal coupling.

A quick inspection shows that Eqs.~\eqref{eq:m_N_N} and~\eqref{eq:m_N_R} are trivially satisfied across all of the parameter space of interest (see Fig.~\ref{fig:neff_BBN}), that Eq.~\eqref{eq:m_H} is redundant with the condition required to avoid thermalizing the scalar sector, that Eq.~\eqref{eq:m_Rho_N} is satisfied by the mixings in the seesaw limit (see Eq.~\eqref{eq:seesaw_DiracY}), an that Eq.~\eqref{eq:m_Rho_R} poses the only relevant requirement, which matches our Eq.~\eqref{eq:condition_ARS} in the main text. 

 Finally, in order to asses the 3rd requirement, we must ensure that processes of the type $N\phi \leftrightarrow N \phi$ are not efficient at $T_{\rm lepto}$ in order to ensure that the CP violating oscillations of sterile neutrinos are coherent. The rate for $N\phi \leftrightarrow N \phi$ processes can be estimated to be $\Gamma  = n_N \left<\sigma v \right> \simeq \lambda_N^4/(576 \pi) \, T \, \xi_N $. Comparing this rate with $H$ we obtain the following bound on the $\lambda_N$ coupling:
\begin{align}\label{eq:condition_ARS_lambdaN_app}
\lambda_N  = \frac{M_N}{v_L}  < 0.07 \,  \sqrt{\frac{T_{\rm lept}}{10^5\,\text{GeV}}} \,  \sqrt{\frac{4\times 10^{-8}}{|h |}} \,,
\end{align}  
This requirement, thus, implies a mild hierarchy between $M_N$ and $v_L$. 

Therefore, if sterile neutrinos, $\rho$'s and $\phi$'s are not populated during reheating, and $|\lambda_{\Phi H}|< 10^{-7}$ and $\Lambda_N \lesssim 0.07$, we can conclude that \textit{i)} the $U(1)_L$ symmetry is spontaneously broken before the onset of sterile neutrino oscillations, and \textit{ii)} $\rho$'s, $\phi$'s and $N$'s do not thermalize with the SM plasma prior to the electroweak phase transition, and \textit{iii)} the CP violating oscillations will be coherent at the time at which the lepton asymmetry is generated. This confirms that ARS leptogenesis can remain viable in the singlet majoron model with symmetry breaking scales  $v_L \lesssim 1\,\text{TeV}$.

Admittedly, we have not carried out a rigorous calculation of leptogenesis within this framework. What our discussion shows is that in the parameter space of interest, our model converges to the conventional models in which ARS leptogenesis has been shown to be successful. It may be possible to further relax some of the identified requirements and still generate the baryon asymmetry of the Universe, however this requires detailed calculations which are beyond the scope of this work. We note that since the temperature of the phase transition can be adjusted, it could be chosen so as to potentially enhance the CP oscillation rate over some finite range of temperatures. Naively, this may enhance the primordial lepton asymmetry and broaden the parameter space in which ARS leptogenesis can be successful.

\section{Primordial Majorons from GeV-scale Sterile Neutrinos}\label{sec:PrimordialMaj_app}

In this appendix, we outline the calculation detailing the creation and decoupling of the primordial majoron population. As discussed in the main text, sterile neutrinos with $0.1\,\text{GeV} \lesssim M_N\lesssim 10\,\text{GeV}$ responsible for generating the active neutrino masses generically thermalize with the SM plasma after the electroweak phase transition~\cite{Ghiglieri:2016xye}. Let us begin with a rough estimation to identify the relevant evolution after sterile neutrino thermalization, and then we will return to a more quantitive assessment.

\begin{figure}
\centering
\includegraphics[width=0.98\textwidth]{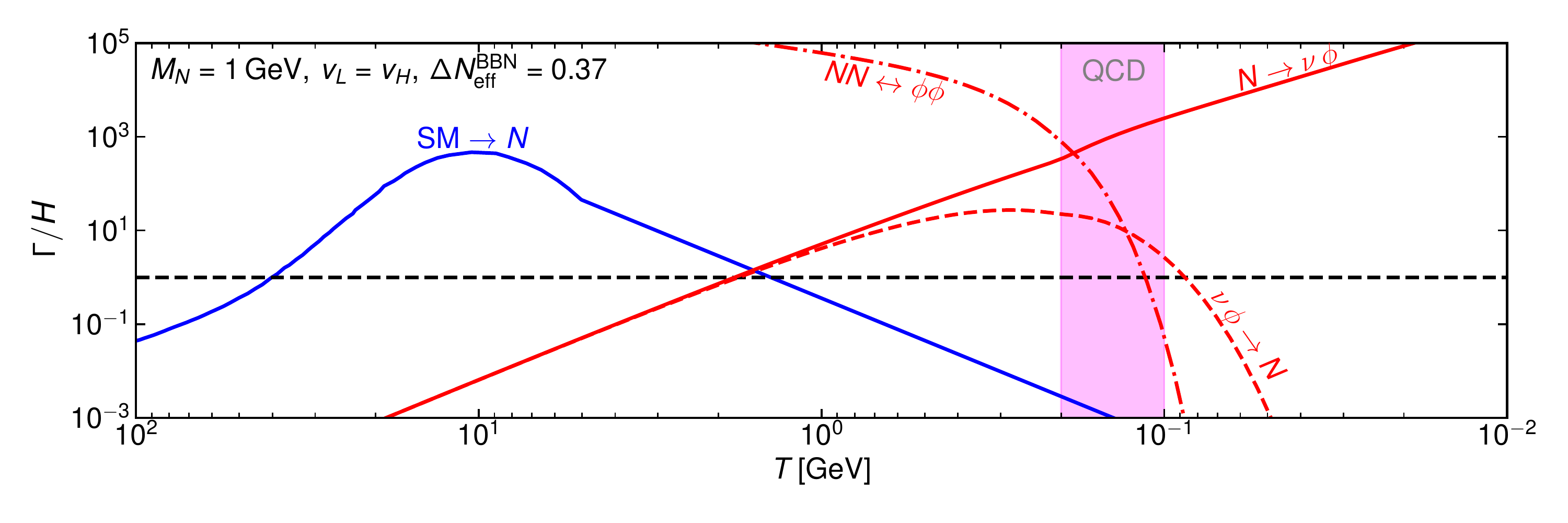}
\vspace{-0.4cm}
	\caption{Relevant rates normalized to the expansion rate for the case of $M_N = 1\,\text{GeV}$ and $v_L = v_H = 246\,\text{GeV}$ which yields $\Delta N_{\rm eff}^{\rm BBN} = 0.37$. In red we show the relevant rates involving majorons and in blue we show the rate of $N$ production from the SM bath~\cite{Ghiglieri:2016xye} (for $T< 5\,\text{GeV}$ we have extrapolated it like $\Gamma \sim T^5$ as relevant for electroweak interactions). For this particular case we can clearly appreciate that annihilation interactions are highly efficient, although inverse decays $\nu\,\phi \to N$ are the last process to decouple at $T\sim 100\,\text{MeV}$.  }\label{fig:N_rates} 
\end{figure}

At high temperatures sterile neutrinos annihilate efficiently to majorons, and will quickly generate a thermal population of them. The annihilation cross section ($N N \leftrightarrow \phi \phi$) responsible for thermalizing majorons at high temperatures is given by
\begin{align}\label{eq:cross_sec}
\sigma_{N \bar{N} \to \phi \phi}(s) = \frac{M_N^2  \sqrt{1-\frac{4 M_N^2}{s}}-2 \frac{M_N^4}{s} \log \left[\frac{s \sqrt{1-\frac{4 M_N^2}{s}}-2 M_N^2+s}{2 M_N^2}\right]}{64 \pi  v_L^4 \left(1-\frac{4 M_N^2}{s}\right)}\,,
\end{align}
where we have consider the limit $m_\rho \gg M_N$ for simplicity. By comparing the rate  $\Gamma \sim n_N \left<\sigma v\right>$ with $H$, one can see that at temperatures near the sterile neutrino mass, $\Gamma(\bar{N}N \leftrightarrow \phi \phi)/H \sim 300\,(M_N/\text{GeV})^2 \left(2\,\text{TeV}/v_L\right)^4 $. This implies thermal equilibrium will be achieved if $v_L< 8\,\text{TeV} \left(\frac{M_N}{\rm GeV}\right)^{3/4}$, which is valid in all  of the parameter space of interest. At lower temperatures this rate will fall below the rate for sterile neutrino decay $N\to \phi\,\nu$. The sterile neutrino abundance will be strongly depleted by  $T \lesssim M_N/10$ (since inverse decays are no longer efficient in restoring the population), and thus the primordial majoron population will decouple near this epoch. We can estimate the approximate energy density stored in the majoron population at the time of BBN by identifying the temperature at which  inverse decays $\nu\, \phi \to N$ are no longer efficient in altering the majoron distribution function. Since the typical energy exchanged by this process is $E \sim M_N$, this will roughly occur when $(M_N n_N) \Gamma_N \lesssim  \rho_\phi H$.  For $T\ll M_N$ one can write $n_N \simeq g_N e^{-M_N/T}  M_N^2 T \sqrt{T/M_N} /(2\sqrt{2}\pi^{3/2})$, which yields an estimate of the decoupling temperature given by:
\begin{align}
\frac{M_N}{T_d} &\simeq 13 + \log \left[ \frac{m_\nu}{0.05\,\text{eV}}\sqrt{\frac{30}{g_\star}} \left(\frac{v_H}{v_L}\right)^2\right]+ \frac{9}{2}\log \left[ \frac{M_N/T_d}{13}\right]\,.\label{eq:DecouplingTemperature}
\end{align} 
Thus for $v_L \lesssim v_H$, the decoupling of majorons will occur at ${T_d} \lesssim M_N/13$. Provided that $M_N$ is sufficiently small ($M_N\lesssim 1\,\text{GeV}$),  the decoupling will occur after the QCD phase transition, which implies the a sizable primordial majoron population will be generated.  In order to highlight the relevant phenomenology, in Fig.~\ref{fig:N_rates} we show the rates for each of the relevant processes as a function of temperature for $M_N = 1\,\text{GeV}$ and $v_L = v_H$. We can clearly appreciate that the last rate to drop below Hubble in this example is $\nu\,\phi \to N$ at $T\sim 100\,\text{MeV}$. This correspondingly predicts\footnote{Note that $\Delta N_{\rm eff}^{\rm BBN}\simeq 0.3 \left(17.4/g_{S}(T_{\rm dec})\right)^{4/3}$, where we have normalized the expression to $g_{S}(100\,\text{MeV}) \simeq 17.4$~\cite{Laine:2015kra}. } a value of $\Delta N_{\rm eff}^{\rm BBN}\simeq 0.34$, which is in excellent agreement with the result from our computation (described below), which gives $\Delta N_{\rm eff }^{\rm BBN} = 0.37$.

In order to quantitatively study this evolution of the majoron and sterile neutrino populations in more detail, we again have chosen to model the thermodynamic evolution following the formalism developed in~\cite{Escudero:2018mvt,Escudero:2020dfa}. This formalism assumes that sterile neutrinos and majorons are both described by thermal equilibrium distributions with evolving temperatures and chemical potentials. This is a good approximation in this context of this problem because: \textit{i)} majorons are massless, \textit{ii)} sterile neutrinos have thermal abundances, and \textit{iii)} sterile neutrinos start to decay while relativistic provided that $v_L < 2\,\text{TeV}$. The last requirement can be clearly seen by comparing the decay rate with the Hubble at $T = M_N/3$:
\begin{align}
\frac{\Gamma(N\to \nu \phi)}{H(T=M_N/3)} \simeq  2 \left(\frac{2\,\text{TeV}}{v_L}\right)^2 \left(\frac{m_\nu}{0.05\,\text{eV}}\right) \sqrt{\frac{50}{g_\star}}\,,
\end{align}
where we have normalized the number with respect to $g_\star \sim 50$, as would be  relevant for $T \sim 200\,\text{MeV}$.

In this framework, the time evolution of the temperature and chemical potentials of the relevant species reads:
\begin{subequations}\label{eq:full_system_maj}
\begin{align}
\frac{dT_\phi}{dt} &=\frac{1}{\frac{\partial n_\phi}{\partial \mu_\phi} \frac{\partial \rho_\phi}{\partial T_\phi}-\frac{\partial n_\phi}{\partial T_\phi} \frac{\partial \rho_\phi}{\partial \mu_\phi} }\left[ -3 H  \left((p_\phi+\rho_\phi)\frac{\partial n_\phi}{\partial \mu_\phi}-n_\phi \frac{\partial \rho_\phi}{\partial \mu_\phi} \right)+ \frac{\partial n_\phi}{\partial \mu_\phi}  \frac{\delta \rho_\phi}{\delta t} - \frac{\partial \rho_\phi}{\partial \mu_\phi}  \frac{\delta n_\phi}{\delta t} \right] , 
 \\
\frac{d\mu_\phi}{dt} &=\frac{-1}{\frac{\partial n_\phi}{\partial \mu_\phi} \frac{\partial \rho_\phi}{\partial T_\phi}-\frac{\partial n_\phi}{\partial T_\phi} \frac{\partial \rho_\phi}{\partial \mu_\phi} } \left[ -3 H \left((p_\phi+\rho_\phi)\frac{\partial n_\phi}{\partial T_\phi}-n_\phi \frac{\partial \rho_\phi}{\partial T_\phi} \right)+ \frac{\partial n_\phi}{\partial T_\phi}  \frac{\delta \rho_\phi}{\delta t} - \frac{\partial \rho_\phi}{\partial T_\phi} \frac{\delta n_\phi}{\delta t} \right]   , \\
\frac{dT_N}{dt} &=\frac{1}{\frac{\partial n_N}{\partial \mu_N} \frac{\partial \rho_N}{\partial T_N}-\frac{\partial n_N}{\partial T_N} \frac{\partial \rho_N}{\partial \mu_N} }\left[ -3 H  \left((p_N+\rho_N)\frac{\partial n_N}{\partial \mu_N}-n_N \frac{\partial \rho_N}{\partial \mu_N} \right)+ \frac{\partial n_N}{\partial \mu_N}  \frac{\delta \rho_N}{\delta t} -  \frac{\partial \rho_N}{\partial \mu_N}  \frac{\delta n_N}{\delta t} \right] , 
 \\
\frac{d\mu_N}{dt} &=\frac{-1}{\frac{\partial n_N}{\partial \mu_N} \frac{\partial \rho_N}{\partial T_N}-\frac{\partial n_N}{\partial T_N} \frac{\partial \rho_N}{\partial \mu_N} } \left[ -3 H \left((p_N+\rho_N)\frac{\partial n_N}{\partial T_N}-n_N \frac{\partial \rho_N}{\partial T_N} \right)+ \frac{\partial n_N}{\partial T_N}  \frac{\delta \rho_N}{\delta t} - \frac{\partial \rho_N}{\partial T_N} \frac{\delta n_N}{\delta t} \right]  ,  \\
\frac{dT}{dt} &=  \left[ - 3H (\rho_{\rm SM}+p_{\rm SM}) + \frac{\delta \rho_\nu}{\delta t} \right]/c_{\rm SM}(T) \,,
\end{align}
\end{subequations}
where in these expressions $\rho_i$, $n_i$, and $p_i$ are the energy density, number density, and pressure of the given species $i$. In addition, $c_{\rm SM} \equiv d\rho_{\rm SM}/dT$ is the heat capacity of the SM plasma that, together with $\rho_{\rm SM}$ and $p_{\rm SM}$, we take from~\cite{Laine:2015kra}. 

In the Maxwell-Boltzmann approximation, we have analytic expressions for the energy and number density exchange rates for decay $N\leftrightarrow \nu \,\phi$ processes which read:
\begin{align}
\left.\frac{\delta n_\nu}{\delta t} \right|_{\rm dec} &= \frac{\Gamma_N  M_N^2 }{2 \pi ^2}\left[ e^{\frac{\mu_N}{T_N}}T_N K_2\left(\frac{M_N}{T_N}\right)- e^{\frac{\mu_\phi}{T_\phi}}  \sqrt{T T_\phi}    K_2\left(\frac{M_N}{\sqrt{T T_\phi}}\right)\right]\,,\\
\left. \frac{\delta \rho_\phi}{\delta t} \right|_{\rm dec} &= \frac{\Gamma_N  M_N^3 }{2 \pi ^2}\left[ e^{\frac{\mu_N}{T_N}}T_N K_2\left(\frac{M_N}{T_N}\right)- e^{\frac{\mu_\phi}{T_\phi}}  T_\phi K_2\left(\frac{M_N}{\sqrt{T T_\phi}}\right)\right]\,,\label{eq:Majoronenergytransfer} \\
\left. \frac{\delta \rho_\nu}{\delta t}  \right|_{\rm dec} &= \frac{\Gamma_N  M_N^3 }{2 \pi ^2}\left[ e^{\frac{\mu_N}{T_N}}T_N K_2\left(\frac{M_N}{T_N}\right)- e^{\frac{\mu_\phi}{T_\phi}}  T   K_2\left(\frac{M_N}{\sqrt{T T_\phi}}\right)\right]\,,
\end{align}
where energy and number density conservation in the decay process implies that:
\begin{align}
\left. \frac{\delta \rho_N}{\delta t}  \right|_{\rm dec}  &= -\left. \frac{\delta \rho_\nu}{\delta t}  \right|_{\rm dec} - \left.\frac{\delta \rho_\phi}{\delta t}  \right|_{\rm dec} \,,\\
\left. \frac{\delta n_\nu}{\delta t}\right|_{\rm dec}  &= \left. \frac{\delta n_\phi}{\delta t}\right|_{\rm dec}  = - \left. \frac{\delta n_N}{\delta t}\right|_{\rm dec} \,.
\end{align}

In addition, in the Maxwell-Boltzmann approximation, the rates for annihilations $\bar{N}N \leftrightarrow \phi\phi$ given Eq.~\eqref{eq:cross_sec} read:
\begin{align}
 \left. \frac{\delta \rho_N}{\delta t}  \right|_{\rm ann}   &= - \left.\frac{\delta \rho_\phi}{\delta t}  \right|_{\rm ann} =  \frac{3 M_N^2 }{32 \pi ^5 v_L^4} \left[T_\phi^7 e^{\frac{2 \mu_\phi }{T_\phi}}e^{-\frac{M_N^2}{8 T_\phi^2}}-T_N^7 e^{\frac{2 \mu_N }{T_N}}e^{-\frac{M_N^2}{8 T_N^2}}\right] \,,\\
\left. \frac{\delta n_N}{\delta t}\right|_{\rm ann}  &= - \left. \frac{\delta n_\phi}{\delta t}\right|_{\rm ann} = \frac{M_N^2}{32 \pi ^5 v_L^4} \left[ T_\phi^6 e^{\frac{2 \mu_\phi }{T_\phi}}e^{-\frac{M_N^2}{6 T_\phi^2}}-T_N^6 e^{\frac{2 \mu_N }{T_N}}e^{-\frac{M_N^2}{6 T_N^2}} \right] \,.
\end{align}
where for the sake of simplicity we have taken $\sigma(s) \simeq {M_N^2}/(64 \pi  v_L^4) $. This is  a good approximation for all temperatures of interest.

The final ingredients needed to obtain an estimation of the size of the primordial majoron population are the initial conditions. Given that \textit{i)} sterile neutrinos thermalize after the electroweak phase transition with the SM plasma, and that \textit{ii)} annihilations between sterile neutrinos and majorons are highly efficient across our relevant parameter space, we start with initial conditions corresponding to all species in thermal equilibrium. In particular, we use:
\begin{align}
T_\gamma = T_N = T_\phi = 100 \,M_N\,,\qquad \mu_N = \mu_\phi = 10^{-2} \times T_{N} \,,
\end{align}
with $t_0 = 1/(2H(T_\gamma))$. Note that for numerical stability we choose these numerical values for the chemical potentials, but that the results are equivalent to choosing $\mu_N = \mu_\phi = 0$ since $n\propto e^{\mu/T}$ and therefore a ratio $1:100$ has a negligible impact on any relevant thermodynamic quantity. The results of solving these equations are shown in Fig.~\ref{fig:neff_BBN} where we display $\Delta N_{\rm eff}^{\rm BBN}$ as a function of $M_N$ and $v_L$.

\section{Big Bang Nucleosynthesis}\label{app:bbn} 

The requirement of successful BBN yields relevant constraints on primordial populations of majorons as parametrized by $\Delta N_{\rm eff}$. At present, the two primordial abundances used to constrain non-standard expansion histories at the time of nucleosynthesis are~\cite{pdg,Allahverdi:2020bys}: Helium-4, $Y_p$, and Deuterium, ${\rm D/H}$. On the one hand, $Y_p$ is very sensitive to the expansion history of the Universe because its abundance is mainly controlled by the time at which deuterium starts to form, which corresponds to $T_D \simeq 0.073\,\text{MeV}$~\cite{Mukhanov:2003xs}. In addition, $Y_P$ is only logarithmically sensitive to the baryon abundance, $\Omega_b h^2$, and its prediction has a negligible theoretical uncertainty. On the other hand, the deuterium abundance is strongly dependent upon the baryon energy density while only moderately dependent upon the expansion rate, potentially modified by $\Delta N_{\rm eff}$. Importantly, although the recent results from the LUNA collaboration~\cite{Mossa:2020gjc} have reduced the theoretical prediction for deuterium, it is still at the $\sim 2.8\%$ level~\cite{Pisanti:2020efz} -- see also~\cite{Pitrou:2020etk} and~\cite{Yeh:2020mgl} which report 1.5\% and 4.4\% uncertainties, respectively. In order to understand the effect of these constraints in our parameter space we have used the predictions and theoretical uncertainties from the recent analysis of~\cite{Pisanti:2020efz}. We contrast these predictions to the measured deuterium abundance~\cite{Cooke:2017cwo}: ${\rm D/H} = (2.527\pm 0.030)\times 10^{-5}$, and choose to do an analysis for two values of $Y_P$. One from the recent analysis of~\cite{Aver:2020fon} that yields $Y_P = 0.2453\pm 0.0034$, and the one from~\cite{Izotov:2014fga} that yields $Y_P = 0.2551 \pm 0.0022$. The reason we choose these two values is because although most recent determinations of $Y_P$ agree within error bars with~\cite{Aver:2020fon}, the determination of $Y_P$ is far from trivial and could have systematic uncertainties. Therefore, we consider the two to highlight the size of potential systematic uncertainties. This being said, we believe that it is likely that the real value is closer to that of~\cite{Aver:2020fon}. 

\begin{figure}
	\includegraphics[width=0.48\textwidth]{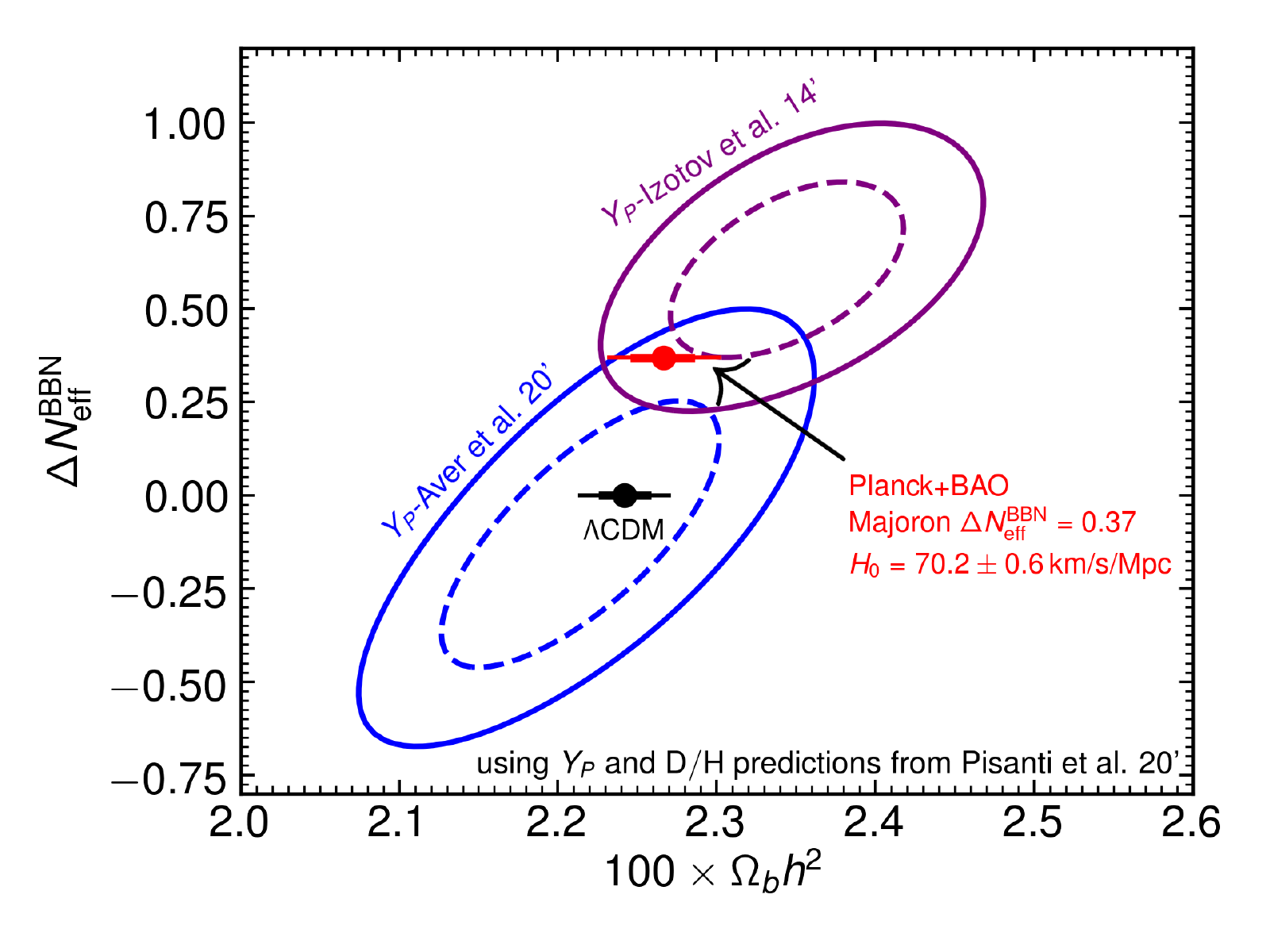}
	\vspace{-0.6cm}
	\caption{BBN constraints on $\Delta N_{\rm eff}^{\rm BBN}$ as a function of $\Omega_b h^2$. Contours correspond to $1$ and $2\sigma$ CL. We have calculated the predictions using the results of~\cite{Pisanti:2020efz}, the deuterium measurements from~\cite{Cooke:2017cwo}, and performed two analyses for the helium measurements of~\cite{Aver:2020fon} and~\cite{Izotov:2014fga}. In red we show the $1$ and $2\sigma$ CL posterior from our CMB analyses in Section~\ref{sec:cosmo}. We can appreciate that our region of interest is in agreement within $2\sigma$ with successful BBN irrespectively of the adopted $Y_P$  value. }\label{fig:BBN} 
\end{figure}

In Fig.~\ref{fig:BBN} we show the resulting constraints from successful BBN on $\Delta N_{\rm eff}$ as a function of $\Omega_bh^2$ when considering the determination of $Y_p$ from~\cite{Aver:2020fon} (blue) and~\cite{Izotov:2014fga} (purple). In red we show the 1-2$\sigma$ CL posterior from our Planck+BAO analysis within our benchmark majoron cosmology with a fixed $\Delta N_{\rm eff}^{\rm BBN} = 0.37$ (see Section~\ref{sec:cosmo}). We can clearly appreciate that the region of parameter space in our benchmark is compatible within $2\sigma$ with both values of $Y_P$ from~\cite{Aver:2020fon} and~\cite{Izotov:2014fga}. We note that the fact that $\Omega_bh^2$ within the majoron cosmology is shifted upwards with respect to $\Lambda$CDM is relevant since it leads to a better agreement with the measured deuterium abundance. Finally, from this figure we can appreciate that given that $\Omega_bh^2$ cannot be too different from the $\Lambda$CDM value, a very conservative constraint within our cosmology would be $\Delta N_{\rm eff}^{\rm BBN}<0.7$.

\section{Cosmological Evolution of Majorons After BBN}\label{app:majorons_CMB}

As in the previous section, we evolve the majoron and active neutrino distributions after BBN using the formalism of~\cite{Escudero:2018mvt,Escudero:2020dfa}. This procedure has previously been explicitly shown to very accurately reproduce the evolution of majorons which thermalize with neutrinos after they have decoupled from the plasma. The time evolution of the temperature and chemical potential for an arbitrary species is given by
\begin{align}
\frac{dT}{dt} &= \frac{1}{(\partial_\mu n) \, (\partial_T \rho) - (\partial_T n) \, (\partial_\mu \rho)} \left[ -3 \,H \, \left((p+\rho) \partial_\mu n - n \, \partial_\mu \rho \right)+ (\partial_\mu n) \, (\partial_t \rho)   - (\partial_\mu \rho) \, (\partial_t n)  \right]
 \, ,\label{eq:dT_dt_simple} \\
 \frac{d\mu}{dt} &= \frac{-1}{(\partial_\mu n) \, (\partial_T \rho) - (\partial_T n) \, (\partial_\mu \rho)}  \left[ -3 \,H \, \left((p +\rho) \, \partial_T n - n \, \partial_T \rho  \right) + (\partial_T n) \, (\partial_t \rho)  - (\partial_T \rho) \, (\partial_t n) \right]   \, , \label{eq:dmu_dt_simple} 
\end{align}
where $\rho$, $n$, and $p$ are the energy density, number density, and pressure of species $i$, and the notation $\partial_X$ represents the partial derivative with respect to either the chemical potential, temperature, or time. For $2\leftrightarrow 1$ processes in the Maxwell-Boltzmann limit, the rate of change in the energy and number density of the majoron are given by
\begin{align}
\partial_t n  &= 3\frac{\Gamma_\phi   m_\phi^2 }{2 \pi ^2}\left[T_\nu e^{\frac{2 \mu_\nu }{T_\nu}} K_1\left(\frac{m_\phi}{T_\nu}\right)-T_\phi e^{\frac{\mu_\phi}{T_\phi}} K_1\left(\frac{m_\phi}{T_\phi}\right)\right]\,,\\ 
\partial_t \rho &= 3\frac{\Gamma_\phi  m_\phi^3 }{2 \pi ^2}\left[T_\nu e^{\frac{2 \mu_\nu }{T_\nu}} K_2\left(\frac{m_\phi}{T_\nu}\right)-T_\phi e^{\frac{\mu_\phi}{T_\phi}} K_2\left(\frac{m_\phi}{T_\phi}\right)\right]\,.
\end{align}
The neutrino equivalent is given by $(\partial_t n)_\phi = - 2 (\partial_t n)_\nu$ and $(\partial_t \rho)_\phi = - (\partial_t \rho)_\nu$. We improve upon authors' previous work~\cite{Escudero:2019gvw} by including matter in the evolution of the background, which is relevant for low majoron masses which thermalize near matter-radiation equality. The initial temperature of the majoron fluid is related to the temperature at decoupling, or $\Delta N_{\rm eff}^{\rm BBN}$, and can be related to the photon temperature via
\begin{align}
\frac{T_{\phi}}{T_\gamma} \simeq 0.607 \left[\frac{\Delta N_{\rm eff}^{\rm BBN}}{0.3}\right]^{1/4} \simeq 0.607  \left(\frac{{g_{\star S}^{\rm SM}|_{\rm today}}}{3.93}\right)^{1/3} \,  \left(\frac{17}{g_{\star S}^{\rm SM}|_{\text{dec}}}\right)^{1/3}   \,.
\end{align}  
where we have normalized the expressions to $T_d \simeq 100\,\text{MeV}$.  We take the initial majoron chemical potential to be zero, and evolve the system from temperatures $T_\gamma = 100 \times m_\phi$ until the majorons have effectively decayed and contribute only negligibly to the energy density of the Universe.  

At the level of the perturbations, we adopt two simplifying approximations: \textit{i)} we treat the interacting neutrino+majoron population as a single massless fluid. This is a good approximation because even though majorons eventually become non-relativistic, their change to the equation of state of the system is always small, $<13\%$ (and typically much less than $5\%$) for values of $\Gamma_{\rm eff} > 0.1$, and  \textit{ii)} we take the collision term to be approximately given by the relaxation time approximation~\cite{Hannestad:2000gt}. The latter approximation is equivalent to say that isotropy in a fluid is achieved at a rate $\Gamma = \frac{\delta \rho}{\delta t} \frac{1}{\rho}$. This approach has been shown to be accurate for scenarios with $2\leftrightarrow 2$ scatterings, see~\cite{Oldengott:2017fhy}, however it is in general expected to overestimate the suppression of free-streaming when the typical angles subtended by the interacting particles are small, see~\cite{Chacko:2003dt,Hannestad:2005ex,Barenboim:2020vrr}. This can be understood by considering the fact that many interactions may be required to alter the directionality of particles if the directional change induced from the interaction is small. For ultra-relativistic majorons, the relaxation approximation is thus not expected to hold (because boosted decays maintain directionality). However, for  $\Gamma_{\rm eff} < 10^2$ majoron interactions only become efficient for temperatures $T <  3\,m_\phi$ (see Fig. 2 of~\cite{Escudero:2020dfa}), and thus no significant boost is expected. To be concrete, $\gamma < 9$ at $T < 3\,m_\phi$ and $\gamma < 1.4$ for $T < m_\phi/3$ (corresponding to the point at which the rate is maximum). Thus, even though we do not calculate the full collision term, we expect it to be a very reasonable description.

The above simplifications allow us to write the density $\delta$, the velocity $\theta$, the shear $\sigma$, and the higher anisotropic moments of the phase space distribution in the synchronous gauge as~\cite{Ma:1995ey}:
\begin{subequations}\label{eq:Hierarchy}
\begin{align}
\dot{\delta}_{\nu\phi} &= -\frac{4}{3}\theta_{\nu\phi} - \frac{2}{3}\dot{h} \,, \\
\dot{\theta}_{\nu\phi} &= k^2 \left(\frac{1}{4}\delta_{\nu\phi} -\sigma_{\nu\phi} \right)  \,,  \\
\dot{F}_{\nu\phi} {}_{2} &= 2\dot{\sigma}_{\nu\phi} = \frac{8}{15}\theta_{\nu\phi}-\frac{3}{5}kF_{\nu\phi \,3}  +\frac{4}{15}\dot{h}+\frac{8}{5}\dot{\eta}  - 2\, a\, \Gamma {\sigma}_{\nu\phi}   \,,  \\
\dot{F}_{\nu\phi\,\ell} &= \frac{k}{2\ell + 1} \left[ \ell \, {F}_{\nu\phi \,(\ell-1)} - (\ell +1){F}_{\nu\phi \,(\ell+1)}   \right] - a \,  \Gamma \, {F}_{\nu\phi\,\ell}  \, \hspace{.6cm} {\rm for} \hspace{.3cm}  \ell \geq 3 \, .
\end{align}
\end{subequations}
Here, derivatives are taken with respect to conformal time, $h$ and $\eta$ represent for the metric perturbations, $k$ is defines the given Fourier mode, ${F}_{\nu\phi\,\ell}$ represents the $\ell^{\rm th}$ multipole, $a$ is the scale factor, and  
\begin{align}
\Gamma = \frac{1}{\rho_\nu} \frac{\delta \rho_\nu}{\delta t} = \Gamma_\phi e^{\frac{\mu_\nu}{T_\nu}} \left(\frac{m_\phi}{T_\nu}\right)^3 K_2\left(\frac{m_\phi}{T_\nu}\right)\,,
\end{align}
where $\Gamma_{\phi}$ is the decay width at rest of the majoron, $K_2$ is a modified Bessel function of order 2, and since chemical potentials are small we effectively approximate $ e^{\frac{\mu_\nu}{T_\nu}} \simeq 1$.

\begin{figure*}
\includegraphics[width=\textwidth]{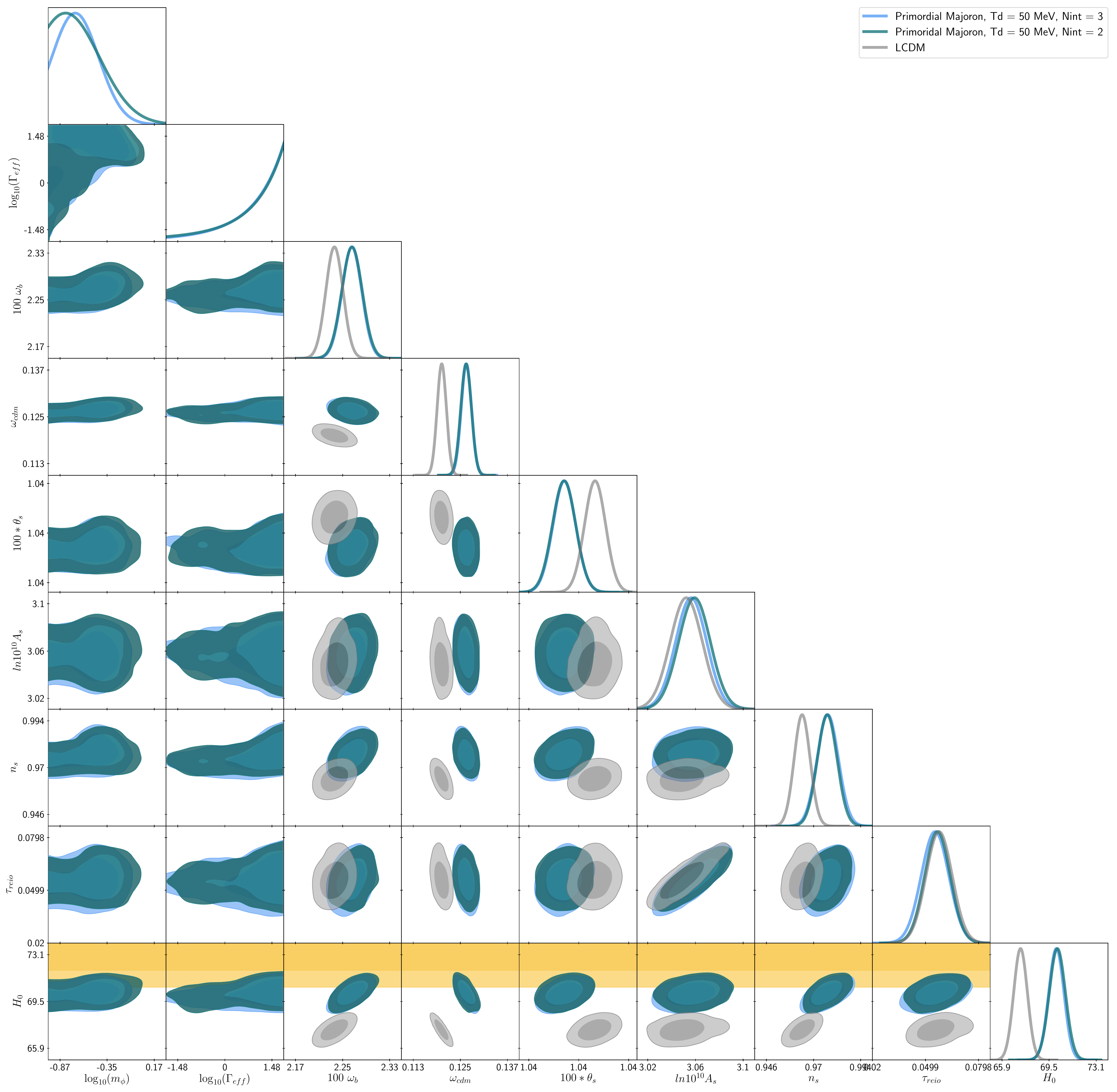}
	\caption{\label{fig:tri_1} Posterior probabilities from an MCMC using {\tt Planck2018+BAO} data, comparing the results of $\Lambda$CDM to the leptogenesis-inspired majoron model with a decoupling temperature $T_d = 50$ MeV and $N_{\rm int} = 2$ (purple), or $N_{\rm int} = 3$ (blue). Results for $N_{\rm int} = 1$ are comparable, and are not shown for clarity.}
\end{figure*}

\begin{figure*}
\includegraphics[width=\textwidth]{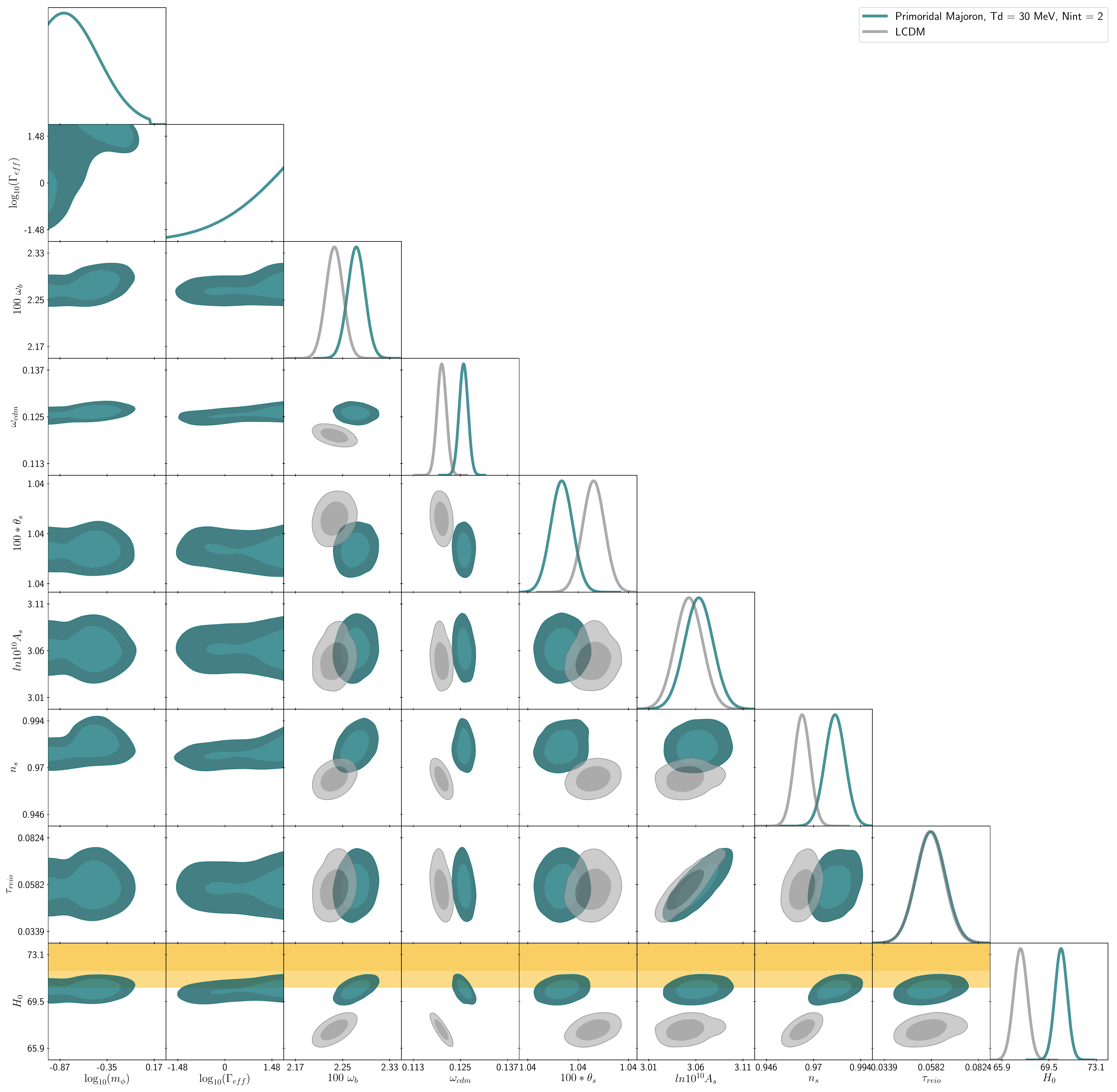}
	\caption{\label{fig:tri_2} Same as \Fig{fig:tri_1} but for the low decoupling temperature model, with $T_d = 30$ MeV, and assuming $N_{\rm int} = 2$. }
\end{figure*}

We implement these perturbations and the background evolution in the Boltzmann code {\tt CLASS}~\cite{Blas:2011rf,Lesgourgues:2011re}, and we run an MCMC using  {\tt Montepython}~\cite{Brinckmann:2018cvx} using {\tt Planck2018+BAO} data on the leptogenesis inspired models with $N_{\rm int} =1, 2, 3$ and $T_d = 50, 30$ MeV (which correspond to $\Delta N_{\rm eff}^{\rm BBN} = 0.37,\,0.48$, respectively). Chains are run until fully converged and all Gelman-Rubin coefficients are $\leq 0.08$. We present here triangle plots in the full parameter space (including derived parameters such as $\sigma_8$); the fiducial model, \ie the one in which we take $T_d = 50$ MeV to be fully consistent with all constraints from BBN, is shown in \Fig{fig:tri_1}, and the result for $N_{\rm int} =3$ and $T_d = 30$ MeV is shown in \Fig{fig:tri_2}. Both plots contain one and two sigma contours from the SH$_0$ES measurement in grey. For completeness, we also include a table (\Tab{table:param_values}) describing the best-fit values and the one sigma uncertainties.

\vspace{0.3cm}

\begin{table}[h]
  \begin{tabular}{l|c|c|c|c}
    \hline\hline
    Parameter            				 & $\Lambda$CDM 	    &  Majoron &  Majoron &  Majoron  \\
     & & $N_{\rm int} = 2$ & $N_{\rm int} = 3$ & $N_{\rm int} = 2$ \\
     & & $T_d = 50 \, {\rm MeV}$ & $T_d = 50 \, {\rm MeV}$& $T_d = 30 \, {\rm MeV}$\\     \hline
     $m_\phi/\text{eV}  $  			    &$-$     	& (0.35)  & (0.31) &  (0.30)  \\
    $\Gamma_{\rm eff} $    			& $-$ 				      			&   (67.61) 					& (59.91) 						& (677.92) \\
    $100\,\Omega_b h^2$  			& 2.235 	(2.2197)  $\pm$	0.015 	&  2.267  (2.2700)  $\pm$ 0.017  	& 2.264 (2.2671)  $\pm $ 0.017 & 2.272 (2.2650) $\pm$ 0.016 \\
    $\Omega_{\rm cdm} h^2$		& 1.200 (0.1210)  $\pm$  0.0011		&   0.1265   (0.1267)  $\pm$  0.0014 & 0.1264  (0.1264)  $\pm$ 0.0013 & 	0.1259 (0.1266) $\pm$ 0.0012 \\
    $100~\theta_s$         			&  1.0419 (1.0420)  $\pm$	  0.0003  		& 1.0411   (1.0410)  $\pm$ 0.0003 	& 1.0410 (1.0412)  $\pm$ 0.0003 &	1.0411 (1.0410)   $\pm$ 0.0003\\
    $\ln(10^{10}A_{s }) $   			 & 3.044	(3.0343)  $\pm$	0.014 	&3.0587   (3.0572)  $\pm$  0.0154 	&  3.056 (3.0536)   $\pm$ 0.015 &	 3.061 (3.0594)   $\pm$ 0.015 \\
    $n_s$               				 &	0.962  (0.9614)  $\pm$	 0.004 	&0.9767    (0.9790)  $\pm$  0.0051 	&  0.977   (0.9781)  $\pm$ 0.005 &		 0.981 (0.9884)   $\pm$ 0.006 \\
    $\tau_{\rm reio}$         			& 0.056    (0.0509)  $\pm$	0.007  		& 0.0562   (0.0576)  $\pm$  0.0078 	& 0.055  (0.0554)  $\pm$ 0.008  & 	0.573 (0.0604)  $\pm$  0.008 \\
    \hline
    $H_0$  [km/s/Mpc]               				 &  67.31  	(66.9165)  $\pm$	   0.53 &  70.18   (70.2969)  $\pm$ 0.61 & 70.06   (69.9915)  $\pm$ 0.60 & 70.15 (70.1460)  $\pm$ 0.48 \\
    \hline
    $(R-1)_{\rm min}$  			&      	0.013				&     0.078	& 0.080 &   0.0233   \\
    $ \chi^2_{\rm min} $ high-$\ell$  	&       2342.76	   	&  2341.78   & 2341.93 &  		2348.40       \\
    $ \chi^2_{\rm min} $ lowl	&      24.00	&    22.35    & 22.48 &		   21.25    \\
    $ \chi^2_{\rm min} $ lowE  	&     396.00 &    396.64  &396.19 & 	397.26    \\
    $ \chi^2_{\rm min} $ lensing  	&      8.91      	&   9.148 & 9.193 &   		9.278       \\
    $ \chi^2_{\rm min} $ BAO  	&       7.57	&             4.905 & 4.839 &   	4.923	       \\
        $ \chi^2_{\rm min} $ CMB  &      2771.67	&   2769.79   & 2769.76 &		   2776.1    \\
        $ \chi^2_{\rm min} $ TOT  	&        2779.24			&   2774.7   & 2774.6 &		 2781.1       \\\hline
$\chi^2_{\rm min}-\chi^2_{\rm min}|^{\Lambda {\rm CDM}}$  		&   0	& -4.54  & -4.64  &    1.86 		   \\
    \hline\hline

  \end{tabular}
  \caption{Mean (best-fit) values with $\pm 1\sigma$ errors of the cosmological 
parameters reconstructed from our combined analysis of 
Planck2018+BAO data in each scenario. Note that the $2\sigma$ lower limit on $\Gamma_{\rm eff}$ for each model (from left to right) corresponds to 0.089, 0.113, and 0.154.}
  \label{table:param_values}
\end{table}

\end{document}